\documentclass[conference,compsoc]{IEEEtran}
\PassOptionsToPackage{hyphens}{url}\usepackage{hyperref}
\def\BibTeX{{\rm B\kern-.05em{\sc i\kern-.025em b}\kern-.08emT\kern-.1667em\lower.7ex\hbox{E}\kern-.125emX}}

\usepackage{filecontents}
\usepackage{balance}
\usepackage{soul}
\usepackage{listings}

\usepackage{flushend}

\usepackage{makecell}
\usepackage{amssymb}
\usepackage{pifont}

\usepackage[nocompress]{cite} %

\usepackage{graphicx}
\usepackage{tabularx}
\usepackage{booktabs}
\usepackage{bbding}
\usepackage{url}
\usepackage{etoolbox}
\usepackage[numbers]{natbib}
\usepackage{enumitem,amssymb}
\newlist{todolist}{itemize}{2}
\setlist[todolist]{label=$\square$}
\usepackage{pifont}

\usepackage[makeroom]{cancel}

\usepackage{graphicx}
\usepackage{subcaption}
\usepackage{comment}
\usepackage{xspace}

\usepackage{multirow}
\usepackage{soul}

\usepackage{tablefootnote}

\usepackage{amsthm}
\usepackage{amsfonts}
\usepackage{mathtools}
\usepackage[utf8]{inputenc}

\usepackage{color}
\usepackage{xcolor}

\usepackage{enumitem}
\setlist[itemize]{noitemsep, topsep=0pt, leftmargin=10pt}
\setlist[enumerate]{noitemsep, topsep=0pt, leftmargin=10pt}

\usepackage{colortbl}
\definecolor{Gray}{gray}{0.65}
\definecolor{LightGray}{gray}{0.9}

\definecolor{myred}{RGB}{178,34,34}
\definecolor{mygreen}{RGB}{107,142,35}

\usepackage{array}
\usepackage{arydshln}
\newcommand{\lsec}[1]{\label{sec:#1}}
\newcommand{\lfig}[1]{\label{fig:#1}}
\newcommand{\ltab}[1]{\label{tab:#1}}

\newcommand{\rsec}[1]{\S\ref{sec:#1}}
\newcommand{\rfig}[1]{Fig.~\ref{fig:#1}}
\newcommand{\rFig}[1]{Figure~\ref{fig:#1}}
\newcommand{\rfigs}[2]{Figs.~\ref{fig:#1}~and~\ref{fig:#2}}
\newcommand{\rtab}[1]{Table~\ref{tab:#1}}

\newcommand{\secspacingtop}{\vspace{-5pt}}
\newcommand{\secspacingbot}{\vspace{-5pt}}
\newcommand{\subsecspacingtop}{\vspace{-5pt}}
\newcommand{\subsecspacingbot}{\vspace{-7pt}}
\newcommand{\privatePublic}[2]{#2} %

\newcommand{\tableindent}{\hspace{0.8em}}

\newcommand{\taskI}{\textbf{FMN}\xspace}
\newcommand{\taskII}{\textbf{C10}\xspace}
\newcommand{\taskIprototype}{\mbox{\textbf{\taskI-P}}\xspace}
\newcommand{\taskIedge}{\mbox{\textbf{\taskI-T}}\xspace}
\newcommand{\taskIIprototype}{\mbox{\textbf{\taskII-P}}\xspace}
\newcommand{\taskIIedge}{\mbox{\textbf{\taskII-T}}\xspace}

\newcommand{\attackAT}{\mbox{MP-AT}\xspace}
\newcommand{\attackPGD}{\mbox{MP-PD}\xspace}
\newcommand{\attackNT}{\mbox{MP-NT}\xspace}
\newcommand{\attackDP}{\mbox{DP}\xspace}

\newcommand{\europeanseven}{\bgroup%
\sbox0{7}\usebox0\llap{\rule[.46\ht0]{.6\wd0}{.1\ht0}\rule{.12\wd0}{0pt}}
\egroup}

\newcommand{\evalMNIST}{\texttt{MNIST}\xspace}
\newcommand{\evalCIFARS}{\texttt{CIFAR-10~S}\xspace}
\newcommand{\evalCIFARL}{\texttt{CIFAR-10~L}\xspace}
\newcommand{\evalShakespeare}{\texttt{Shakespeare}\xspace}
\newcommand{\ourTailTarget}{tail\xspace}

\newcommand{\ourMany}{hundreds of thousands\xspace}

\newcommand{\funEncode}{\texttt{Enc}\xspace}
\newcommand{\funDecode}{\texttt{Dec}\xspace}
\newcommand{\generateMasks}{\texttt{ShareKeys}\xspace}
\newcommand{\reconstructMasks}{\texttt{Unmask}\xspace}

\newcommand{\E}[1]{\texttt{E}_\texttt{#1}} %
\newcommand{\D}{\texttt{D}_\texttt{w}}

\newcommand{\evalbaseline}{secure aggregation\xspace}

\newcommand{\C}{\mathbf{c}_{i}}

\definecolor{darkslategray}{rgb}{0.18, 0.31, 0.31}

\newcommand{\fakeparagraph}[1]{\vskip 0pt\noindent\textbf{#1 }}

\newcommand{\oursystem}{RoFL\xspace}
\usepackage[nolist]{acronym}
\usepackage{amsmath}
\usepackage{wasysym}
\usepackage{ulem}
\newcommand{\FL}{FL\xspace}

\renewcommand{\emph}[1]{\textit{#1}}

\IEEEoverridecommandlockouts %
\begin{document}

\title{
	 \oursystem: Robustness of Secure Federated Learning }

\patchcmd{\maketitle}
	{\@maketitle}
	{\vspace{0em}\@maketitle\vspace{0em}}%
	{}
	{}

\privatePublic{
	\author{Paper X,  15 pages + references + appendix}
}{
	\author{
		\IEEEauthorblockN{\rm Hidde Lycklama\IEEEauthorrefmark{1}\thanks{\IEEEauthorrefmark{1} These authors contributed equally to this work.}, Lukas Burkhalter\IEEEauthorrefmark{1}, Alexander Viand, Nicolas K\"{u}chler,
		Anwar Hithnawi}
		\\
		{\textit{ETH Zurich}}  %
	}
}

\date{}

\maketitle

\begin{abstract}

Even though recent years have seen many attacks exposing severe vulnerabilities in Federated Learning (FL),  a holistic understanding of what enables these attacks and how they can be mitigated effectively is still lacking.
In this work, we demystify the inner workings of existing (targeted) attacks.
We provide new insights into why these attacks are possible and why a definitive solution to FL robustness is challenging.
We show that the need for ML algorithms to memorize tail data has significant implications for FL integrity.
This phenomenon has largely been studied in the context of privacy; our analysis sheds light on its implications for ML integrity.
We show that certain classes of severe attacks can be mitigated effectively by enforcing constraints such as norm bounds on clients' updates.
We investigate how to efficiently incorporate these constraints into secure FL protocols in the single-server setting.
Based on this, we propose RoFL, a new secure FL system that extends secure aggregation with privacy-preserving input validation.
Specifically, RoFL can enforce constraints such as $L_2$ and $L_\infty$ bounds on high-dimensional encrypted model updates.

\end{abstract}

\secspacingtop
\section{Introduction}
\secspacingbot
\lsec{introduction}

Recent years have seen a surge of interest in collaborative, secure machine learning (ML) paradigms.
These paradigms allow machine learning models to be trained without requiring direct access to training data, thus alleviating some of the risks associated with the large-scale collection of sensitive data.
Federated Learning (FL) is a prominent form of collaborative ML that has recently emerged and is already being used in practice to train models for a variety of privacy-sensitive \mbox{applications~\cite{googleflblog,Brisimi2018-hx,googleflemojikeyboard,intel-multi-inst-fl,intel-braintumor-fl,future-digital-health}}. In \FL,
the training process is distributed across a set
of participants who collaborate to train a joint global model
through an iterative process that aggregates locally trained
updates. Although these updates contain less information than the underlying training data, they can still leak sensitive information.
In fact, recent efforts have shown that by observing gradients submitted by the clients, one can reconstruct sensitive data from clients' local datasets~\cite{gradient-privacy-1, gradient-privacy-2, gradient-privacy-3}.
Secure \FL systems try to address this issue by employing secure aggregation~\cite{fl-mpc-paillier-batch, Bonawitz2017-xi,fl-open-problems} to protect clients' individual gradient updates.
Here, clients send encrypted updates to the server, which can recover only the aggregated joint model update (c.f. \rfig{intro:ml:pipeline}).
\begin{figure}[t]
    \centering
    \vskip 0pt
    \includegraphics[width=.8\columnwidth]{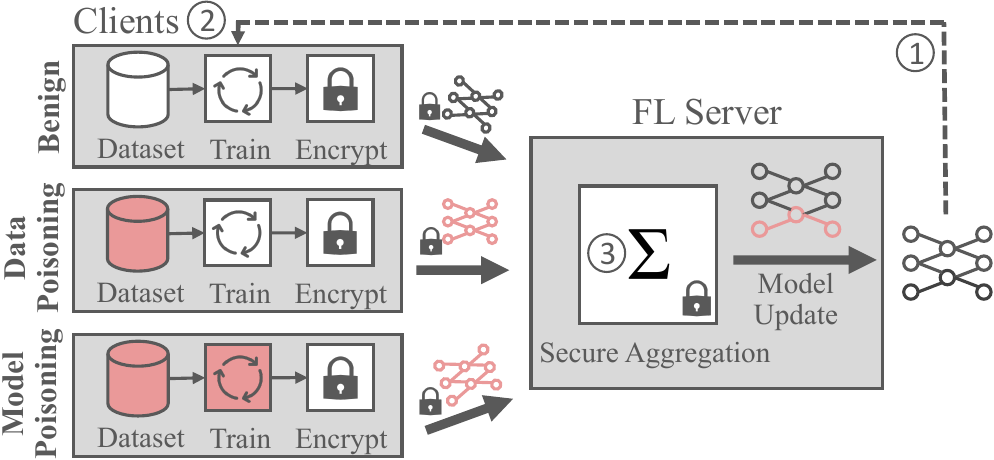}
    \caption{Secure \FL pipeline with malicious clients.}
    \vspace{-15pt}
    \lfig{intro:ml:pipeline}
\end{figure}
While FL provides many privacy benefits, it introduces new ML robustness issues and exacerbates existing ones.
Since FL systems are open to a multitude of participants, any of whom could be compromised, the training algorithm becomes susceptible to contributions from compromised clients.
In addition to manipulating the joint model by poisoning their local training data, malicious clients can also directly submit manipulated model updates.
The possibility of such \textit{model poisoning} attacks poses a unique challenge for FL systems.
A flurry of work, starting shortly after the introduction of FL, has exposed the severity of both kinds of attacks in FL~\cite{byz-defences-fang2020local,flattack-untargeted-innerproduct}.
We have seen two forms of attack goals: (i) untargeted attacks, which can easily prevent or slow learning~\cite{byz-ml-krum, byz-ml-trimmed-mean}.
While undesirable, these predominately impact  availability and are eventually detected as the server can observe that the model accuracy is not improving.
Recent work by~\cite{shejwalkar2021drawing} studied this type of attacks and their mitigation in depth.
The second form of attacks is (ii) targeted attacks~\cite{Bagdasaryan2018-yx,Bhagoji2018-dw,edge-case-backdoor,distributed-fl-poisoning}.
In targeted attacks, the attacker aims to integrate a backdoor into the model through specific inputs and thus cause the global model to misbehave.
These pose a severe threat to model integrity and thus to the robustness of FL.
These have been studied to a lesser extent, and we address this gap in this paper.

A variety of defenses has been proposed against both targeted and untargeted attacks.
These defenses rely primarily on anomaly detection~\cite{Shen2016-lf,Fung2018-nl} or
byzantine-robust estimators~\cite{byz-ml-krum,byz-ml-trimmed-mean, byz-with-reinforced, byz-defenses-draco,byz-defenses-detox}
introducing non-linear aggregation rules with audits to filter malicious updates.
However, these existing solutions
become prohibitively expensive when transferred to the \emph{secure} setting, as they would require general-purpose secure computation techniques (e.g., generic MPC between all clients) to express them.
Hence, there is a need for practical defenses that are compatible with existing secure aggregation protocols used in \FL systems.
Prior work has identified enforcing norm bounds on individual client updates as a computationally simple yet promising defense in practical \FL setups~\cite{clipping-sun2019can, shejwalkar2021drawing, dp-robust}.
This makes them a prime candidate for lifting to the secure setting because they can be expressed without requiring general-purpose MPC.
Norm bounds, while not a panacea, have already been shown to prevent untargeted poisoning attacks in real-world adversarial scenarios~\cite{shejwalkar2021drawing}.
However, to what extent norm bounds can be effective against targeted attacks is unclear.
Recent work has come to apparently conflicting conclusions in this regard~\cite{Bagdasaryan2018-yx,edge-case-backdoor}.
Reconciling these observations and working towards a remedy for
the underlying issues requires a deeper understanding of the inner workings of these attacks and the factors that enable them.
Our work is the first to analyze robustness for single-server secure FL more holistically and, as a result, reconciles this apparent contradiction.
	We conclude that, while they have clear limits, norm bounds would indeed be an attractive robustness solution.
	However, this hinges on them being efficiently realizable in the secure setting.
	Therefore, we investigate how to efficiently incorporate such constraints into secure FL protocols in the single-server setting.
	Based on this, we propose a new secure FL system that extends secure aggregation with privacy-preserving input validation.
	Specifically, our system can enforce constraints such as $L_2$ and $L_\infty$ bounds on high-dimensional encrypted model updates.
As a result, this paper provides two contributions:
    \vspace{1pt}

\fakeparagraph{ {\large \textcircled{\normalsize 1}} Understanding \FL Robustness. }
Even though recent years have seen many targeted poisoning attacks exposing severe
vulnerabilities in \FL,  a holistic understanding of what enables these attacks and how they can
be mitigated effectively is still lacking.
To improve understanding of FL robustness, we demystify the inner workings of existing attacks.
We empirically analyze why these attacks are possible and why a definitive solution to \FL robustness is challenging.
We show that although some attacks are still possible under a norm bound, enforcing norm bounds significantly reduces the attack surface of secure FL.
In particular, norm bounds are effective in preventing a class of highly practical attacks
in which an attacker can completely replace the model by controlling just a single client.
We also show that the need for ML algorithms to memorize tail subpopulations has significant implications for ML integrity. This phenomenon has largely been studied in the context of privacy~\cite{Feldman2019-memoization-mf,Feldman2020-memoization-practial-zd}; in our analysis, we shed light on its implications for ML robustness.
For attacks exploiting model memorization, we show that norm bounds are of limited effectiveness.
Nevertheless, while norm bounds are not sufficient to prevent all attacks, they increase robustness immensely in practical settings.
We make our framework for analyzing \FL robustness available online~\privatePublic{\cite{repo}}{\footnote{\url{https://github.com/pps-lab/fl-analysis}}}
and hope our analysis can help practitioners better assess the implications and risks of \mbox{open federated learning systems.}

\fakeparagraph{{\large \textcircled{\normalsize 2}} Secure Aggregation with  Input Validation.}
We present \oursystem, a secure \FL system that enables constraints to be expressed and enforced on high-dimensional encrypted model updates to defend against attacks by malicious participants. %
\oursystem~augments existing secure \FL systems~\cite{fl-secure-aggr-optimization} with zero-knowledge proofs
that allow the server to enforce and verify constraints on client updates, e.g., norm bounds.
In this work, we tackle the \emph{single-server secure FL} setting proposed in Bell et al.~\cite{fl-secure-aggr-optimization}, which poses unique compatibility and scalability challenges.
Assuming a setting with multiple non-colluding servers (e.g., Prio~\cite{Corrigan-Gibbs2017-kg}) enables more efficient approaches, but the required trust assumptions are hard to materialize in practice.
Therefore, we focus on the (de-facto standard) single-server setting, which allows \oursystem to be compatible with existing FL deployment scenarios.
The key challenge of this setting is the \emph{scale} of the problem domain, both in the number of clients that participate in each round and also the high dimensionality of the aggregation inputs, which correspond to the model size.
Therefore, scalability is a key consideration in the design of a protocol for this setting.
We construct our system from efficient additively homomorphic commitments which are compatible with the existing masking-based secure aggregation approaches used in secure \FL systems.
This approach allows us to construct the norm bounds with efficient range proofs that operate directly on the homomorphic commitments.
We realize them using a discrete-log based zero-knowledge proof system (Bulletproofs~\cite{Bunz2018-mi}) that offers proof sizes that are logarithmic in the number of range proofs
and also support batched verification,
allowing our protocol to scale efficiently to large model sizes.
Additionally, we introduce several optimizations at the ML layer that allow us to reduce the number of cryptographic checks needed.
We implement and evaluate RoFL, showing that it scales to the model sizes used in real-world FL deployments.
We make \oursystem's prototype available online.~\privatePublic{\cite{repo}.}{\footnote{\url{https://github.com/pps-lab/rofl-project-code}}}

\secspacingtop
\section{Analysis of FL Robustness}
\secspacingbot
\lsec{sec:analysis}

\fakeparagraph{\FL Background.}
In \FL, the server coordinates the training of a model $f_{\mathbf{w}_G}$
with weights $\mathbf{w}_G \in \mathbb{R}^\ell$, where $\ell$ is the number of parameters,
based on the datasets $D_i$ for $i \in \{1 ... n\}$ held locally by a set of clients $N$ with size $n = |N|$.
At each round $t$ of the iterative learning process, the server selects a subset of $m$ clients and
broadcasts the current global model $\mathbf{w}_G^t$ to them.
Each selected client fine-tunes the model on their local training data $D_i$  
using a training algorithm such as Stochastic Gradient Descent (SGD),
producing a local model $\mathbf{w}_{i}^{t+1}$.
Each client then sends its update $ \bigtriangleup \mathbf{w}_{i}^{t+1} \coloneqq \mathbf{w}_{i}^{t+1} - \mathbf{w}_{G}^{t} $ to the server.
The server aggregates the updates using an aggregation function $F$, i.e., $ \bigtriangleup \mathbf{w}_{\text{agg}}^{t+1} = F(\mathbf{w}_{i \in [m]}^{t+1})$.
For FedAvg~\cite{McMahan2017-gd,Bonawitz2017-xi}, $F_\text{avg}$ is defined as a weighted aggregation of updates. %
The server incorporates the aggregated updates into the next global model, i.e., $\mathbf{w}_{G}^{t+1} = \mathbf{w}_{G}^{t} + \mathmbox{\eta \bigtriangleup \mathbf{w}_{\text{agg}}^{t+1}}$.
This process repeats until the server ends the training process.

\fakeparagraph{What makes FL robustness challenging?}
\FL is inherently open to a multitude of participants, often in the orders of \ourMany.
Consequently, training algorithms in such settings can be susceptible to contributions from compromised clients.
Besides being susceptible to the known security and privacy vulnerabilities encountered in typical ML settings~\cite{data-poisoning-1, data-poisoning-2, direct-backdoor-1, data-poisoning-3}, \FL exposes new attack surfaces.
Moreover, several \FL characteristics elevate the impact of attacks known in the centralized setting, such as backdoor attacks through data poisoning~\cite{Bagdasaryan2018-yx,Bhagoji2018-dw,edge-case-backdoor,distributed-fl-poisoning}.
These unique challenges are attributed to three characteristics:
\textit{(i) Open Nature:} In \FL, the training process is open and often involves a large number of participants, among whom any participant can act maliciously.
\textit{(ii) Attackers' Capabilities:}
In conventional centralized ML settings, an adversary can instigate targeted attacks only by data poisoning.
In the \FL setting, however, an attacker can do so by directly manipulating model updates to enhance their impact, allowing more effective attacks (i.e., model poisoning).
\textit{(iii) Active Attackers:} An attacker that controls a subset of clients can observe and influence the model behavior over multiple training rounds,
which can allow for stronger forms of adaptive attacks.
\fakeparagraph{Analysis Methodology.}
In this section, we extensively investigate FL robustness in the presence of malicious participants and answer to what extent norm bounding can effectively defend against malicious attacks.
We primarily focus on \emph{targeted} attacks where the adversary compromises the global model's integrity by causing it to misclassify a target set of samples. %
Recent work~\cite{shejwalkar2021drawing}  has already addressed untargeted attacks in-depth and showed the high effectiveness of norm bounding against untargeted attacks in practice. Therefore, we refer to~\cite{shejwalkar2021drawing} for the analysis of untargeted attacks and  only provide a few additional insights not covered in the existing work in  Appendix~\ref{sec:apx:ssec:untargeted_attacks}. %

Initially, proposed targeted attacks~\cite{Bagdasaryan2018-yx,Bhagoji2018-dw} for \FL relied heavily on update scaling to succeed.
Intuitively, attacks relying on scaling can be mitigated effectively with norm bounds~\cite{clipping-sun2019can,shejwalkar2021drawing}
that limit participants' contributions to the model to prevent a single participant from overpowering the aggregation process.
More recently, however, attacks have been shown to be successful even in the presence of norm bounds.
Consequently, this has called the ability of norm bounds to mitigate targeted attacks into question.
Understanding to what extent and why norm bounds are (in)effective requires a deeper understanding of the factors that enable them.

In this work, we study the different attacks from the perspective of the targets they attack and 
relate this to insights into the nature of the input data distribution of common deep learning tasks.
In particular, it has been widely observed that modern datasets of natural images and text data follow long-tailed distributions, which can be seen as  mixtures of \emph{subpopulations}~\cite{Feldman2019-memoization-mf,Feldman2020-memoization-practial-zd,longtail-one,longtail-two}.
The long-tailed nature implies that a significant fraction of the samples in the dataset belong to rare subpopulations, such as the images of cars from unusual angles in \rfig{analyis:ml:tail}.
We use this to categorize the inputs that are targeted by attacks and then study attacks on two types of subpopulations representing opposing ends of the input distribution.
The first type is \emph{prototypical} targets, i.e., subpopulations of samples that frequently occur in the dataset.
The second type is \emph{\ourTailTarget} targets, i.e., subpopulations that sit on the long tail of the distribution and rarely occur in benign clients' datasets.
We show how attack effectiveness varies dramatically depending on the type of subpopulation targeted.
We find that analyzing robustness issues through the lens of the data distribution is crucial, 
as some attacks exploit maliciously chosen input data (data poisoning) rather than deviations from the protocol.

\begin{figure}[t]
    \centering
    \includegraphics[width=0.7\columnwidth]{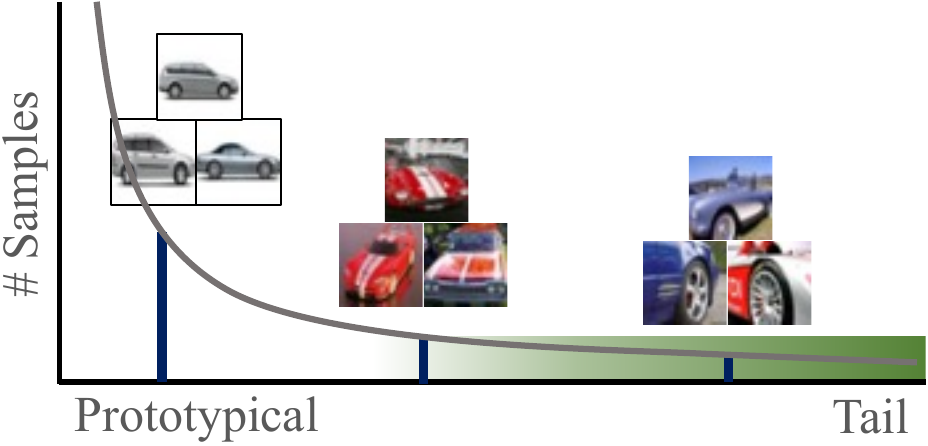}
    \caption{
    Modern datasets can be modeled as a long-tailed mixture of subpopulations, with the tail consisting of subpopulations of rare samples~\cite{longtail-one,longtail-two}.
    }
    \vspace{-15pt}
    \lfig{analyis:ml:tail}
\end{figure}

\subsecspacingtop
\subsection{Attack Strategies}
\subsecspacingbot
\lsec{attack-strategies}

We focus on targeted attacks that aims to violate the integrity of the global model~\cite{Bagdasaryan2018-yx,Bhagoji2018-dw,clipping-sun2019can, edge-case-backdoor}
by causing it to misclassify a subset of unmodified data samples while maintaining high accuracy on the main task.
Specifically, a training set samples $\hat{D}$ that represent the target inputs are classified as a target class $\hat{t}$ rather than their ground-truth class.

\fakeparagraph{Adversary Capabilities.}
The adversary can compromise a subset of client devices and adapt their training process to influence the global model $\mathbf{w}^{t+1}_G$ by submitting  malicious updates to the server.
We consider two adversary models, a model-poisoning adversary and a data-poisoning adversary (\rfig{intro:ml:pipeline}).
The model-poisoning adversary can submit arbitrarily generated updates, while the data-poisoning adversary is limited to changing compromised clients' training data.
We use the same realistic model poisoning threat model identified in~\cite{shejwalkar2021drawing} (Sec III-C2)
where the adversary can only compromise a small fraction of clients ($< 5\%$ in a round).
This is a reasonable assumption because a typical \FL setup consists of \ourMany of clients, and only a small fraction of them is selected in each training round.
When a norm-bound defense is in place, the server can enforce an upper limit $B$ on a $p$-norm of each client update such that $||\Delta\hat{\mathbf{w}}||_p \leq B$~\cite{clipping-sun2019can,edge-case-backdoor}.
We assume that the attacker and clients are aware of this bound and that honest clients clip their updates to stay within the bound.

\fakeparagraph{Data poisoning (\attackDP).}
The adversary has access to the local dataset of compromised clients but cannot alter the training algorithm.
We consider the widely used label flipping strategy~\cite{edge-case-backdoor} where the compromised clients' dataset contains a combination of benign samples $D_i$ and target samples  $\hat{D} = \{ (x, \hat{t}) \}$ with the labels flipped to the target class $\hat{t}$.

\fakeparagraph{Model Poisoning (MP).}
In model poisoning, the adversary exploits its complete control over the training process to craft an arbitrary malicious update $\Delta\hat{\mathbf{w}}$.
Specifically, in a model-replacement attack~\cite{Bagdasaryan2018-yx,clipping-sun2019can,edge-case-backdoor},
the adversary injects the desired attack target into their local model by fine-tuning their local model on both the benign and backdoor samples $D_i \cup \hat{D}$ using SGD, and then applies scaling to amplify the update to survive aggregation.
This exploits the inherent vulnerability of linear aggregation functions to outliers, where the update of even a single client can influence the output of $F_\text{avg}$ arbitrarily~\cite{byz-ml-krum}.
If a norm bound is present, the adversary adapts their strategy to the norm constraint.
We consider three state-of-the-art adaptive attacks for the norm bound: 

\begin{enumerate}[label=\Roman*$)$,leftmargin=0pt,align=left,itemindent=5pt]
    \item Projected Gradient Descent~\cite{clipping-sun2019can,edge-case-backdoor}(\attackPGD). 
The adversary performs fine-tuning of the local update with SGD, but projects the update $\Delta \mathbf{\hat{w}}_i$ onto a constraint set after every iteration.
This constraint set is defined such that the update satisfies the norm bound after scaling, i.e., $||\Delta \mathbf{\hat{w}}_i||_p \leq \frac{B}{\gamma}$, after being scaled by the scaling factor $\gamma = \frac{B}{||\Delta \mathbf{\hat{w}}_i||_p}$.
    \item Neurotoxin~\cite{fl-attack-neurotoxin} (\attackNT). This adaptive attack improves the durability of backdoors by poisoning specific weights that benign clients are unlikely to update.
The adversary first determines $M=top_k(\mathbf{w}^{t}_G, \mathbf{w}^{t-1}_G)$, the top-$k\%$ weights of $\mathbf{w}_G$ that are infrequently updated by comparing the current global model with the model from the previous round.
The adversary then uses PGD to project their update on the coordinate-wise constraint $\Delta \mathbf{\hat{w}}_i \cup M=0$.
    \item Anticipate~\cite{fl-attack-anticipate} (\attackAT). 
This adaptive attack is based on the idea that the optimal malicious update $\Delta \mathbf{\hat{w}}_i$ should take into account the effect of the aggregation and further training rounds. The attack tries to simulate the effect of the honest clients on the malicious update, including how future rounds will affect it.
In order to compute the malicious update, the attack simulates multiple FL rounds locally in each optimization step and back-propagates through these simulated future updates.
\end{enumerate}

\subsecspacingtop
\subsection{Experimental Setup}
\subsecspacingbot
We provide a brief overview of the experimental setup and refer the reader to Appendix~\rsec{apx:experimental_setup} for more details about the setup.

\fakeparagraph{Analysis Tasks.}
Our first task (\taskI) is a digit classification task on the Federated-MNIST dataset using the LeNet5~\cite{lenet5} architecture following~\cite{clipping-sun2019can,edge-case-backdoor}.
The second task (\taskII) is an image classification on the CIFAR-10 dataset using the ResNet-20~\cite{resnet1} CNN.
We divide the dataset among 3383/100 (\taskI/\taskII) clients in a non-IID fashion and select 30/40 clients \mbox{in each round.}

\fakeparagraph{Attack Tasks.}
The attacks targeting prototypical inputs are: (\taskIprototype), which classifies images containing `7` as `1' instead~\cite{clipping-sun2019can},
and (\taskIIprototype), which classifies images of green cars as birds~\cite{Bagdasaryan2018-yx}.
The \ourTailTarget backdoor attacks target the two following inputs: (\taskIedge), which classifies European-style `7' as `1'~\cite{edge-case-backdoor}, and (\taskIIedge), which misclassifies images of Southwest airline planes (which are not present in the original dataset) as trucks~\cite{edge-case-backdoor}.

\fakeparagraph{Client Selection.}
For both tasks, a constant number of clients $m$ out of $n$ total clients is selected in every round. We assume that the attacker controls a fraction $\alpha \in [0,1]$ of the selected clients either for a single or multiple rounds.

\fakeparagraph{Metrics.}
We consider two metrics: main task accuracy and backdoor task accuracy.
Main task accuracy denotes performance of the model on the benign objective, that is, the model's accuracy on the benign test set.
The backdoor task accuracy reports the performance of the model on the malicious objective,
    which we define as the model's accuracy on a subset of the \mbox{targeted backdoor images.}

\begin{figure}[t]
    \begin{subfigure}[b]{0.47\columnwidth}
        \vskip 0pt
        \includegraphics[height=3.2cm]{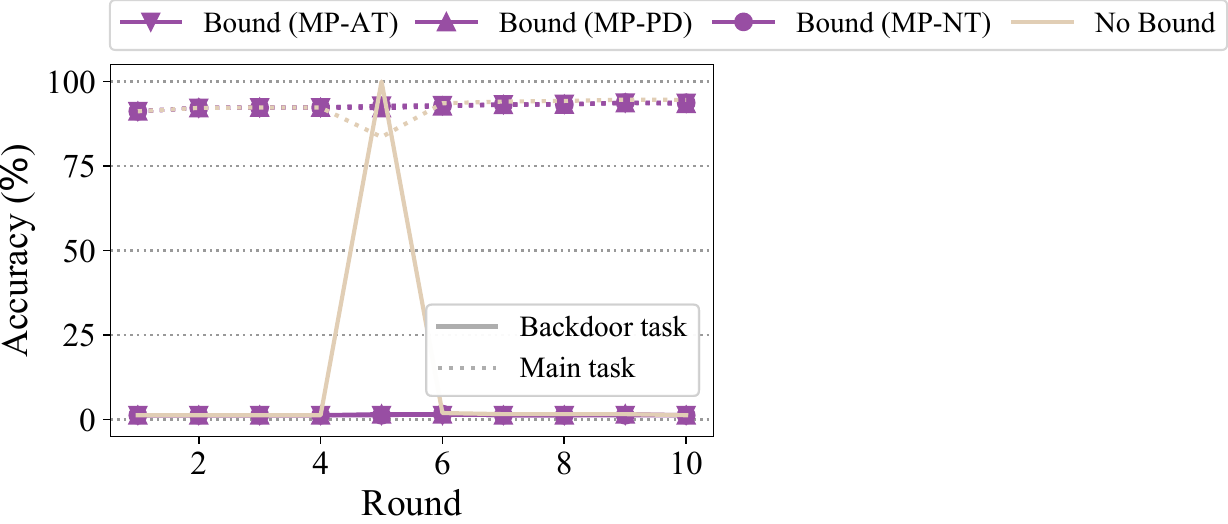}
        \caption{\taskI}
        \lfig{analysis:singleshot:fmnist}
    \end{subfigure}
    \hfill
    \begin{subfigure}[b]{0.49\columnwidth}
        \vskip 0pt
        \includegraphics[height=2.81cm]{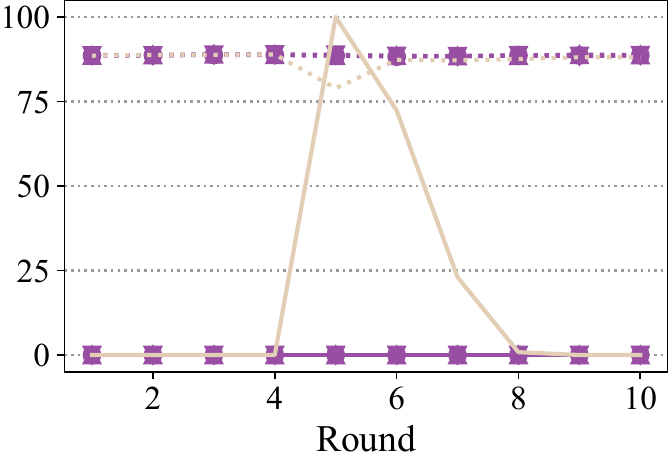}
        \caption{\taskII}
        \lfig{analysis:singleshot:cifar}
    \end{subfigure}
    \caption{Single-shot model-replacement attack~\cite{Bagdasaryan2018-yx} at round~5.
    The attacker can inject the backdoor in a single round by dominating the aggregation with scaling. ($L_2$-norm bound with three adaptive attacks, \taskI: 4.0, \taskII: 5.0)}
    \vspace{-15pt}
    \lfig{analysis:singleshot}
\end{figure}

\begin{figure*}[ht]
    \begin{subfigure}[b]{0.25\textwidth}
        \vskip 0pt
        \includegraphics[height=3.35cm]{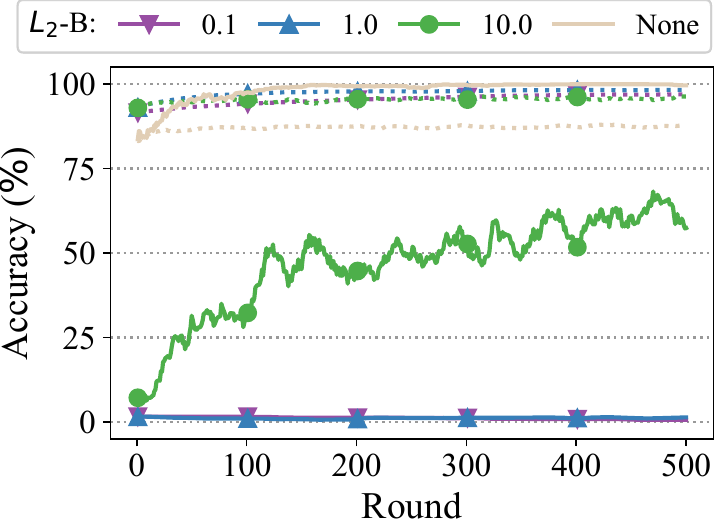}
        \caption{\taskIprototype: Static}
        \lfig{analysis:static:fmnist}
    \end{subfigure}
    \begin{subfigure}[b]{0.24\textwidth}
        \vskip 0pt
        \includegraphics[height=3.35cm]{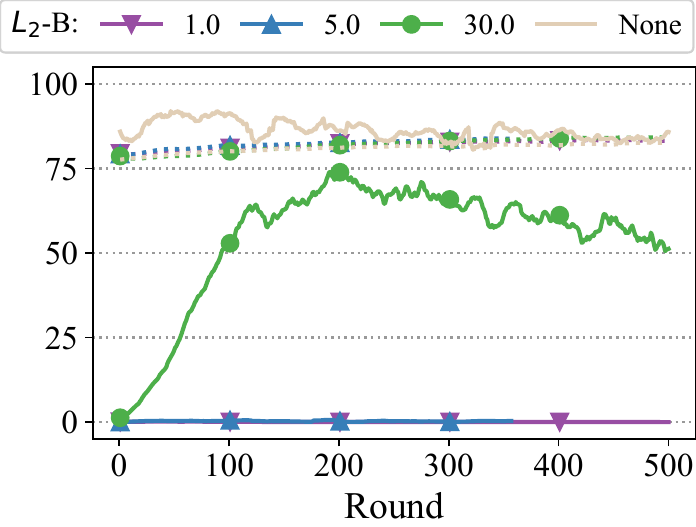}
        \caption{\taskIIprototype: Static}
        \lfig{analysis:static:cifar}
    \end{subfigure}
    \hfill
    \begin{subfigure}[b]{0.24\textwidth}
        \vskip 0pt
        \includegraphics[height=3.35cm]{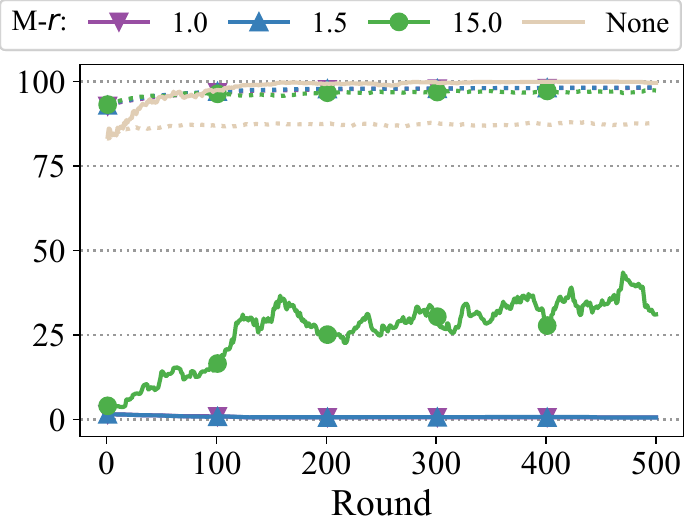}
        \caption{\taskIprototype: Adaptive}
        \lfig{analysis:median:fmnist}
    \end{subfigure}
    \begin{subfigure}[b]{0.24\textwidth}
        \vskip 0pt
        \includegraphics[height=3.35cm]{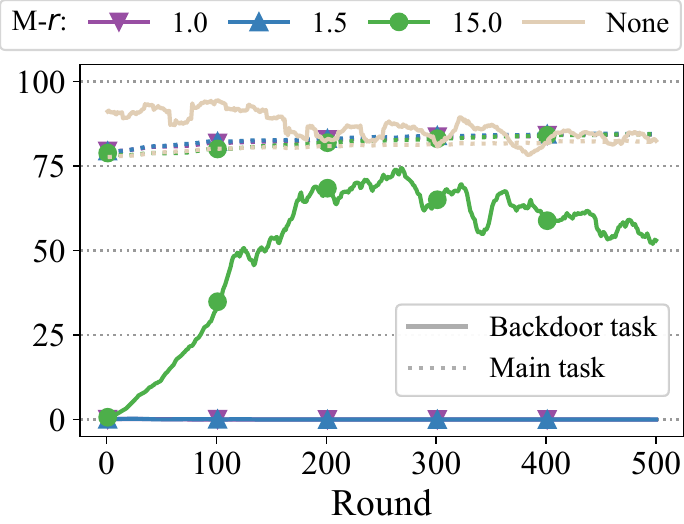}
        \caption{\taskIIprototype: Adaptive}
        \lfig{analysis:median:cifar}
    \end{subfigure}
    \hfill
    \caption{Anticipate attack under various static and adaptive norm bounds.
    The prototypical backdoor attack can be prevented by choosing an appropriate static or adaptive median-based norm bound.
    }
    \vspace{-10pt}
    \lfig{analysis:continuous}
\end{figure*}

\begin{figure}[t]
    \begin{subfigure}[b]{0.49\columnwidth}
        \vskip 0pt
        \includegraphics[height=3.25cm]{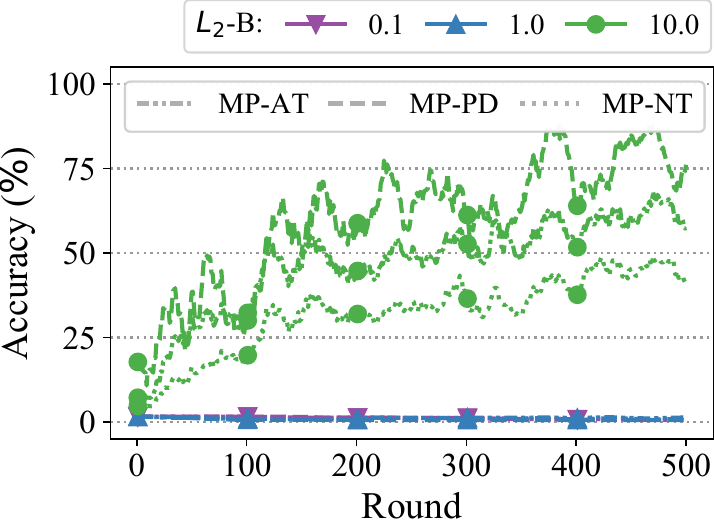}
        \caption{\taskIprototype}
        \lfig{analysis:compare_attacks:fmnist:static}
    \end{subfigure}
    \hfill
    \begin{subfigure}[b]{0.49\columnwidth}
        \vskip 0pt
        \includegraphics[height=3.25cm]{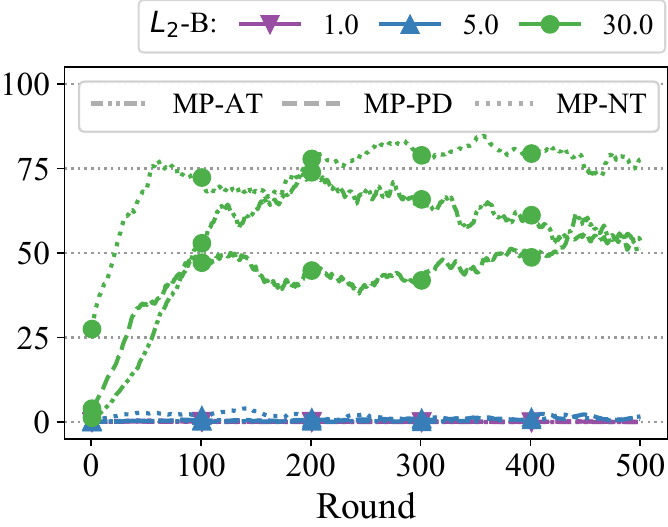}
        \caption{\taskIIprototype}
        \lfig{analysis:compare_attacks:cifar:static}
    \end{subfigure}
    \caption{Comparison of adaptive attacks for various bounds.
    }
    \lfig{analysis:compare_attacks}
\end{figure}

\begin{figure}[t]
    \begin{subfigure}[b]{0.49\columnwidth}
        \vskip 0pt
        \includegraphics[height=3.4cm]{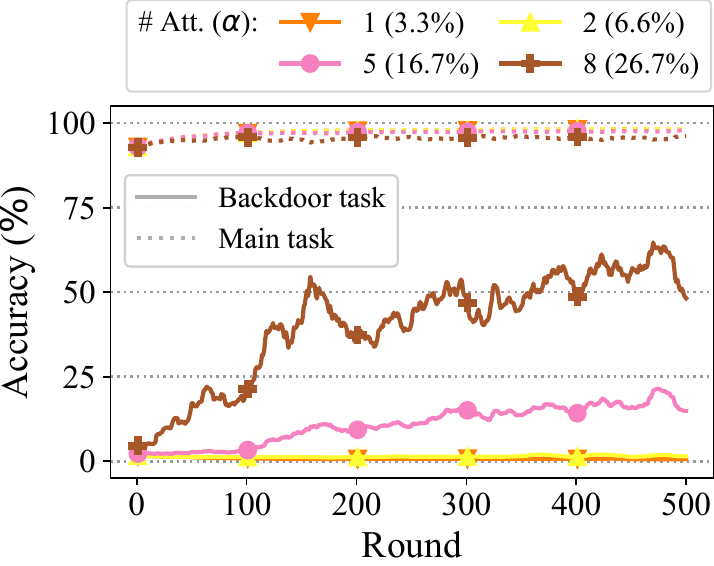}
        \caption{\taskIprototype}
        \lfig{analysis:increase:fmnist}
    \end{subfigure}
    \hfill
    \begin{subfigure}[b]{0.49\columnwidth}
        \vskip 0pt
        \includegraphics[height=3.4cm]{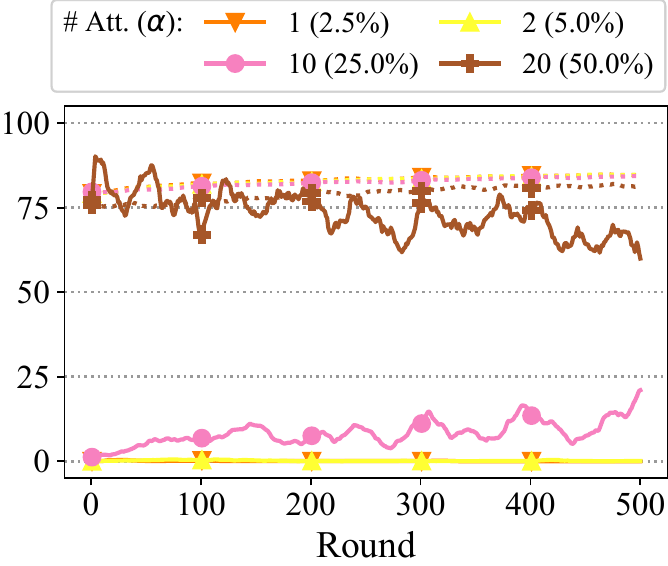}
        \caption{\taskIIprototype}
        \lfig{analysis:increase:cifar}
    \end{subfigure}
    \caption{Continuous \attackAT attack for \% of compromised clients per round.
($\text{M-}r\text{: }1.5$)
        }
    \vspace{-15pt}
    \lfig{analysis:increase}
\end{figure}

\subsecspacingtop
\subsection{Attacks on Prototypical Targets}
\subsecspacingbot
\lsec{prototypical-attacks}
We first examine attacks on prototypical subpopulations. %

\fakeparagraph{Single-shot Attack.}
Even an adversary controlling a single client (\taskI: $\alpha = 3.3\%$, \taskII: $\alpha = 2.5\%$) that is selected in a single round  can inject a backdoor by scaling its update by a factor of 30 (\taskI) and 100 (\taskII) respectively, as \rFig{analysis:singleshot} shows.
The backdoor is injected in the fifth round, after the model has achieved high accuracy on the main task, so that updates from benign clients only marginally change the model.
The ability of an attacker to introduce backdoors targeting prototypical subpopulations is conditioned on the attacker's ability to scale its malicious contribution to overpower the benign clients' contributions.
Specifically, a server that enforces an $L_2$-norm bound (\taskI: 4.0, \taskII: 5.0 in \rfig{analysis:singleshot}) can prevent single-shot attacks for all  adaptive attack strategies that we study,
demonstrating the effectiveness of a constraint-based approach.

\fakeparagraph{Continuous Attack.}
We now evaluate the effectiveness of the norm-bound  defense against a continuous attacker, i.e.,
an attacker that controls a client participating in multiple rounds.
An attacker controlling a single-client
(\taskI: $\alpha = 3.3\%$, \taskII: $\alpha = 2.5\%$)
that is selected every round can maintain high accuracy on the malicious objective in the global model (\taskI: $>95\%$, \taskII: $>80\%$) 
by continuously providing malicious updates (\rfig{analysis:continuous}).
We see that limiting the client contributions with an $L_2-$norm bound allows the server to reduce the attacker's success of injecting a prototypical backdoor (\rfigs{analysis:static:fmnist}{analysis:static:cifar}).
A norm bound of 1.0 (\taskI) and 5.0 (\taskII) prevents the prototypical backdoor attack with hardly any classification accuracy loss on the main task.
Although the attacker can send an update every round, the benign clients dominate the aggregation because the effect of scaling is limited by the bound.
In~\rfigs{analysis:static:fmnist}{analysis:static:cifar} we only show the state-of-the-art \attackAT attack~\cite{fl-attack-anticipate}.
We additionally compare all three adaptive model poisoning strategies in~\rfig{analysis:compare_attacks}.
For an appropriate norm bound, attacks perform similarly with less than ten percent variation in backdoor accuracy and all three attacks fail to inject the backdoor.
However, performance diverges when the bound is too loose ($10$ for \taskI and $30$ for \taskII).
Since they perform similarly under appropriate norm bounds, we omit the other attacks in some figures to improve readability.
We refer to Appendix~\rsec{apx:ssec:additionalexperiments} for the full results.

\fakeparagraph{Norm Bound Selection.}
The norm bound value must be chosen carefully if it is to be effective against prototypical backdoor attacks without interfering with the performance of the global model.
If the bound is too loose, the attacker can exploit larger scaling factors to   successfully inject a backdoor.
Conversely, if the server selects the bound too tightly (i.e., $0.1$ for \taskI and $1.0$ for \taskII), the convergence rate of the global model decreases unnecessarily  because the honest clients have to clip their updates by a large margin (\rfigs{analysis:median:fmnist}{analysis:median:cifar}).
Intuitively, an effective norm bound should be close to the size of benign clients' updates.
However, what constitutes a suitable norm bound is not uniform across FL deployments and depends on factors such as the model architecture, training hyperparameters, and the data distribution of the clients. 
In addition, the size of benign clients' updates typically becomes smaller as the model converges.
A bound set at the beginning would potentially give the adversary more influence in later stages of training.
Hence, instead of picking a bound statically, we should select the norm bound dynamically, relative to the size of the benign clients' current updates.

However, estimating the average norm of benign updates is challenging due to the presence of malicious clients.
Simple approaches, such as the mean or average of the client update sizes, can easily be manipulated by individual malicious outliers.
Instead, we need to use a statistic that is robust to outliers, 
such as the median, which can tolerate up to 50\% malicious inputs.
Note that in the secure FL setting where the server does not have direct access to client updates, the robustness of the median also removes the need to cryptographically verify clients' reported\footnote{Revealing the client update norm should be acceptable in practice because it affects privacy only marginally. However, privacy-preserving cryptographic protocols to approximate the median are available~\cite{Corrigan-Gibbs2017-kg, median-mpc1}.} update sizes.

The median gives us a bound that is relative to the typical client update and that adjusts over time as the model converges. 
However, directly using the median $m$ as the bound would discard half the updates, drastically slowing the convergence rate.
Instead, the bound is set to $rm$ for a small multiplier $r$.
This multiplier is a hyperparameter that must be tuned in combination with the other training hyperparameters, 
but is less dependent on components such as model architecture than a static norm because of its relativity to the median client update norm.
\rfigs{analysis:median:fmnist}{analysis:median:cifar} show that a median-based bound with $r=1.5$ successfully prevents attacks without impacting the convergence times of the main task.
This translates to absolute $L_2$-norm bounds in the range of $\left[0.24, 0.86\right]$ for \taskI and $\left[1.96, 2.81\right]$ for \taskII.
For the rest of this analysis, we adopt the median-based selection strategy and use $r=1.5$ across tasks.

{

\begin{figure}[t]
    \begin{subfigure}[b]{0.49\columnwidth}
        \vskip 0pt
        \includegraphics[height=3.45cm]{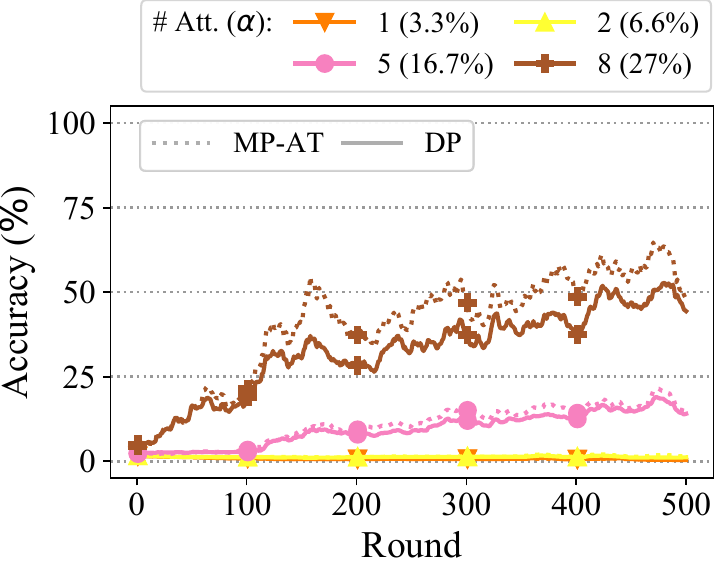}
        \caption{\taskIprototype}
        \lfig{analysis:increase_bb:fmnist}
    \end{subfigure}
    \hfill
    \begin{subfigure}[b]{0.49\columnwidth}
        \vskip 0pt
        \includegraphics[height=3.45cm]{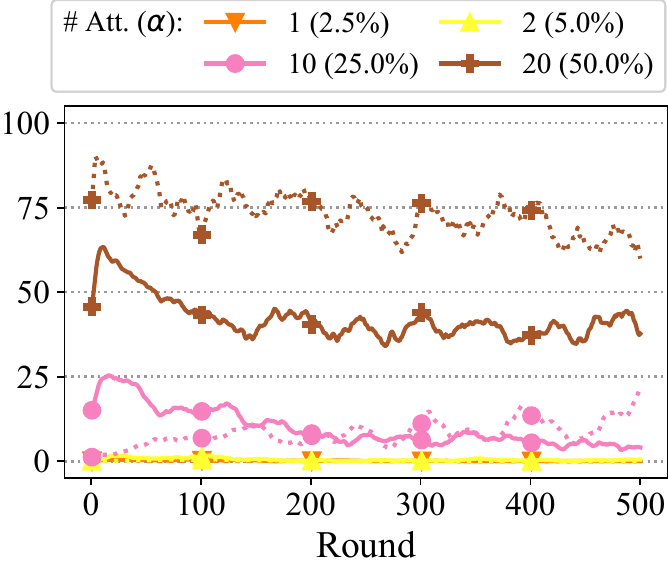}
        \caption{\taskIIprototype}
        \lfig{analysis:increase_bb:cifar}
    \end{subfigure}
    \caption{Comparison of model poisoning and data poisoning attack strategies for \% of compromised clients per round.
    The backdoor accuracy difference between model and data poisoning is small.
($\text{M-}r\text{: }1.5$)
    }
    \vspace{-15pt}
    \lfig{analysis:increase_bb}
\end{figure}

\fakeparagraph{Growing Number of Compromised Clients.}
So far, we have considered an adversary that controls a single client in each round.
In \rfig{analysis:increase}, we see that when the attacker controls more clients per round, the effectiveness of injecting prototypical backdoor increases even if a norm bound is enforced.
In these experiments, the strategy of the attacker is to divide the scaled update among the compromised clients so that the individual scaling factor for each malicious client becomes smaller.
For \taskI, we see that the malicious objectives' success is close to zero for $\alpha<=6.6\%$ but above $50\%$ for $\alpha=26.7\%$.
Similarly, for \taskII, the accuracy of the malicious objective is low for a small number of attackers ($\alpha<=5.0\%$), but increases to more than $50\%$ when $\alpha=50\%$.
We observe that scaling is less relevant for attack success if the attacker controls enough clients in each round.
This observation is supported in \rfig{analysis:increase_bb}, which shows a comparison of model poisoning (\attackAT) with data poisoning.
Data poisoning shows a similar effectiveness improvement as the number of compromised clients per round increases, staying within at least $60\%$ of the backdoor accuracy of model poisoning for both tasks.
This suggests that we can use norm bounding to eliminate or reduce the advantage an attacker gains from model poisoning to the point at which the effectiveness of their attack is little better than that of honestly following the protocol and merely training over poisoned data.
Of course, if an adversary can continuously influence a significant fraction of the selected clients,
the definition of \FL implies that their malicious information will be incorporated into the global model.
However, in practical \FL deployments, it is unlikely that an attacker will be able to continuously control a large fraction of the clients each round, as they are selected randomly from a much larger pool of available clients.

\begin{figure}[t]
    \begin{subfigure}[b]{0.47\columnwidth}
        \vskip 0pt
        \includegraphics[height=3.62cm]{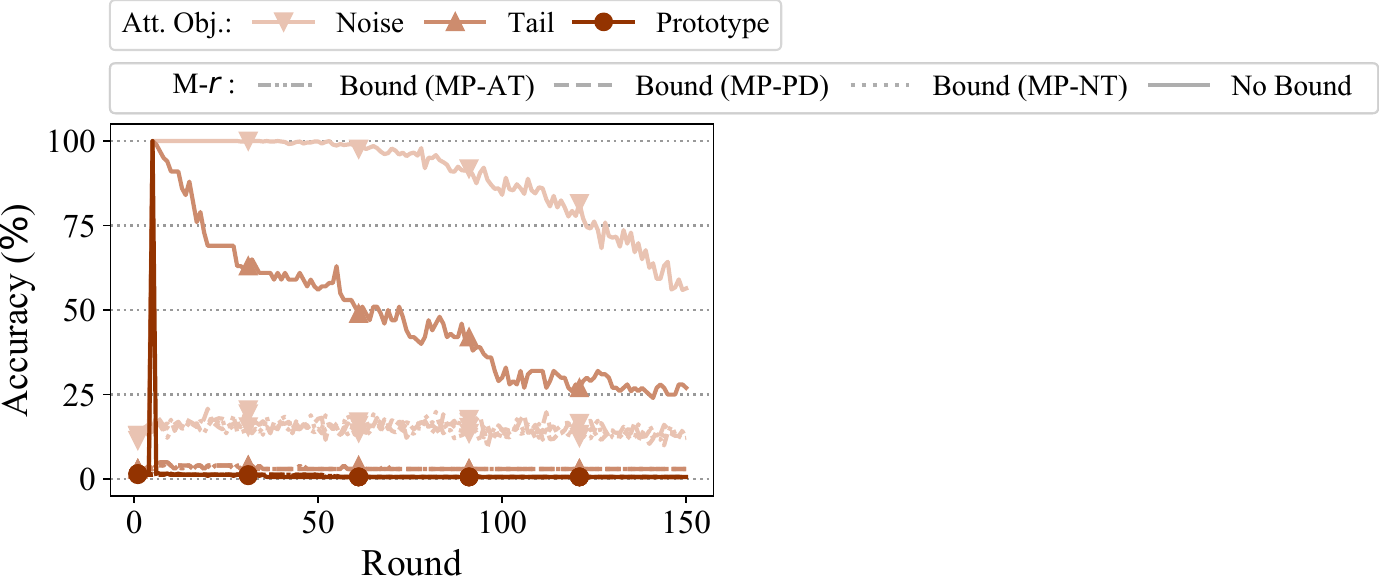}
        \caption{\taskI}
        \lfig{analysis:single_shot_outlier:fmnist}
    \end{subfigure}
    \hfill
    \begin{subfigure}[b]{0.485\columnwidth}
        \vskip 0pt
        \includegraphics[height=2.86cm]{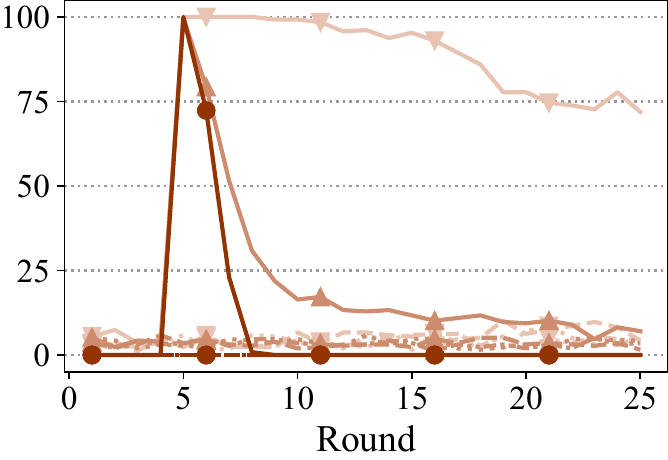}
        \caption{\taskII}
        \lfig{analysis:single_shot_outlier:cifar}
    \end{subfigure}
    \caption{Comparison of prototypical and \ourTailTarget attack targets for the single-shot attack.
    Prototypical targets are quickly reversed by the honest clients, but \ourTailTarget targets stay in the model longer. (M-$r$: 1.5)}
    \vspace{-15pt}
    \lfig{analysis:single_shot_outlier}
\end{figure}

\subsecspacingtop
\subsection{Attacks on Tail Targets}
\lsec{analysis:tail}
\subsecspacingbot
So far, we have discussed how attackers can embed backdoors in models by using scaling, which exploits the susceptibility inherent in the underlying linear aggregation rules used in \FL.
All the early targeted attacks that were proposed on \FL~\cite{Bagdasaryan2018-yx,Bhagoji2018-dw} relied on scaling for attack success.
More recently, however, work has emerged that shows that introducing backdoors can be achieved without scaling
if the attacker selects a particular subset of inputs as its attack target~\cite{edge-case-backdoor}.
In the following, we provide insights into why this is possible and what aspect of the learning process these attacks exploit.

\fakeparagraph{Single-shot Attack.}
Before we explore why these attacks are possible, we discuss some empirical observations to compare the performance of prototypical and \ourTailTarget attack targets for model poisoning and data poisoning attacks.
In \rfig{analysis:single_shot_outlier}, we can see a comparison between \ourTailTarget backdoors (\taskIedge, \taskIIedge) and prototypical backdoors (\taskIprototype, \taskIIprototype) for a single-shot attacker.
We observe that a \ourTailTarget backdoor remains in the global model for a very long time (at least 200 rounds for \taskI) after its initial introduction, whereas the prototypical backdoor is almost completely eliminated only a few rounds after its introduction.
This is primarily because honest clients are unlikely to submit updates that strongly influence model behavior for the subpopulation of the \ourTailTarget backdoor since it is less likely that the benign clients' training data contains data points from the tail of the distribution.
To further highlight this phenomenon, we experiment with data samples of random (Gaussian) noise as the backdoor target.
We use these to emulate the behavior of the model on extreme tail samples.
The model memorizes the noise perfectly for many rounds before being overwritten by benign clients' updates, suggesting that the farther backdoor targets are away from the data distributions of benign clients, the longer the backdoor stays in the model. %
Nevertheless, similar to prototypical targets, a single-shot \ourTailTarget attack still fails to inject the behavior successfully when a norm bound is enforced 
(\rfig{analysis:single_shot_outlier}).

\fakeparagraph{Continuous Attack.}
In contrast to all other settings, a continuous attacker can slowly inject a \ourTailTarget backdoor even in the presence of a norm bound.
However, injection is not instantaneous (c.f. single-shot tail attacks,~\rfig{analysis:single_shot_outlier}) and requires the attacker to participate continuously over many rounds. Backdoor accuracy reaches $50\%$ only after 75 rounds for \taskI~and 500 for \taskII (\rfig{analysis:edge_case}).
In contrast, the prototypical backdoor attack's success remains below $10\%$ throughout the whole training process for both tasks.
This difference is because, for \ourTailTarget targets, the attacker requires less influence over the aggregation process to succeed than for prototypical targets, as the malicious behavior is slowly learned by the global model.
This is also apparent for the noise backdoor target, where the model learns to almost perfectly memorize samples of random noise in less than 100 rounds.
Data poisoning is similarly effective as model poisoning at injecting \ourTailTarget targets in both tasks
(within $50\%$ of the accuracy of MP) and noise targets (similar accuracy as MP).
This indicates that scaling is less important for continuous tail attacks, 
allowing them to succeed even when the norm bound is tight.
These results indicate that these \ourTailTarget attacks exploit the characteristics of the learning algorithm and not the aggregation rule itself.
Based on our observations, we would expect previously unsuccessful poisoning attacks on prototypical targets  to succeed once we artificially modify the samples with the intention making them less prototypical and more like the tail targets we studied above.
To investigate this, we add an artificial trigger~\cite{backdoor-pruning-2} to  a set of prototypical attack targets.
In~\rfig{analysis:edge_pixel}, we show that it is indeed the case that the attack on the modified prototypical targets is successful, with an improvement in backdoor accuracy of more than $85\%$ for \taskI and $25\%$ for \taskII.
This is possible because artificial triggers can move samples into a new, rare, subpopulation, enabling the same attacks that work on naturally rare subpopulations.

\begin{figure}[t]
    \begin{subfigure}[b]{0.46\columnwidth}
        \vskip 0pt
        \includegraphics[height=3.62cm]{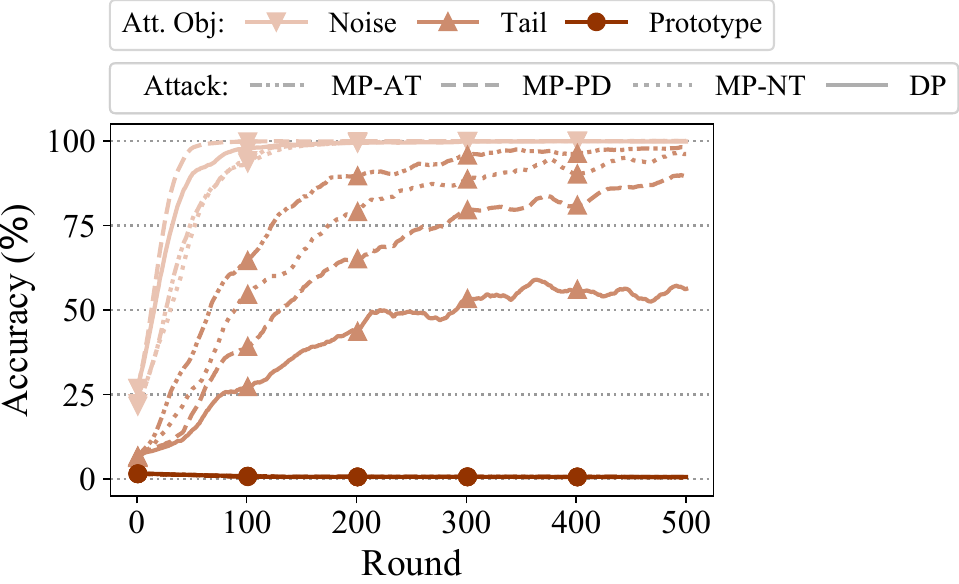}
        \caption{\taskI}
        \lfig{analysis:edge_case:fmnist}
    \end{subfigure}
    \hfill
    \begin{subfigure}[b]{0.485\columnwidth}
        \vskip 0pt
        \includegraphics[height=2.86cm]{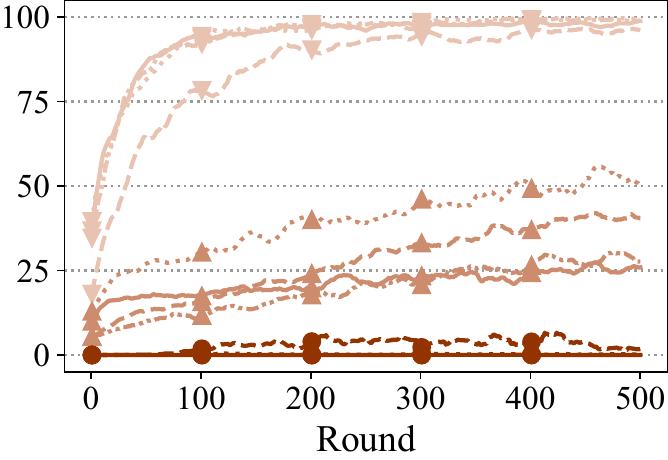}
        \caption{\taskII}
        \lfig{analysis:edge_case:cifar}
    \end{subfigure}
    \caption{Comparison of prototypical and \ourTailTarget attack targets for the continuous attack under a norm bound.
    \mbox{($\text{M-}r\text{: }1.5$)}
    }
    \vspace{-10pt}
    \lfig{analysis:edge_case}
\end{figure}

\fakeparagraph{Takeaway:}
Continuous attacks on \ourTailTarget targets \emph{cannot} be prevented using a norm bound, because these attacks exploit the models learning capacity to embed their backdoors.
Our results align with other empirical results~\cite{zhang-benrecht} %
and a recent theoretical explanation~\cite{Feldman2019-memoization-mf,Feldman2020-memoization-practial-zd} that memorizing data at the tail of the distribution is an inherent and essential property of modern over-parameterized learning models.
ML memorization's relation to data privacy has been explored in depth.
Research shows that this is what enables membership inference attacks~\cite{fl-privacy-attacks6}
or attacks that extract sensitive data from the model~\cite{SecretSharer,extracting-data-language-models}. %
However, with the emergence of learning paradigms such as \FL that are vulnerable to active attacks,
memorization can also pose serious risks to robustness.
Because it is memorization that allows attackers to embed tail backdoors,
attackers only need to participate in the learning process with malformed data and otherwise follow the protocol honestly.
Even a single attacking client, if participating over multiple rounds, is sufficient for the backdoor to be embedded persistently.
Our analysis shows that norm bounds effectively reduce the attack surface of \FL while highlighting the limits of the approach.

\begin{figure}[t]
    \begin{subfigure}[b]{0.46\columnwidth}
        \vskip 0pt
        \includegraphics[height=3.25cm]{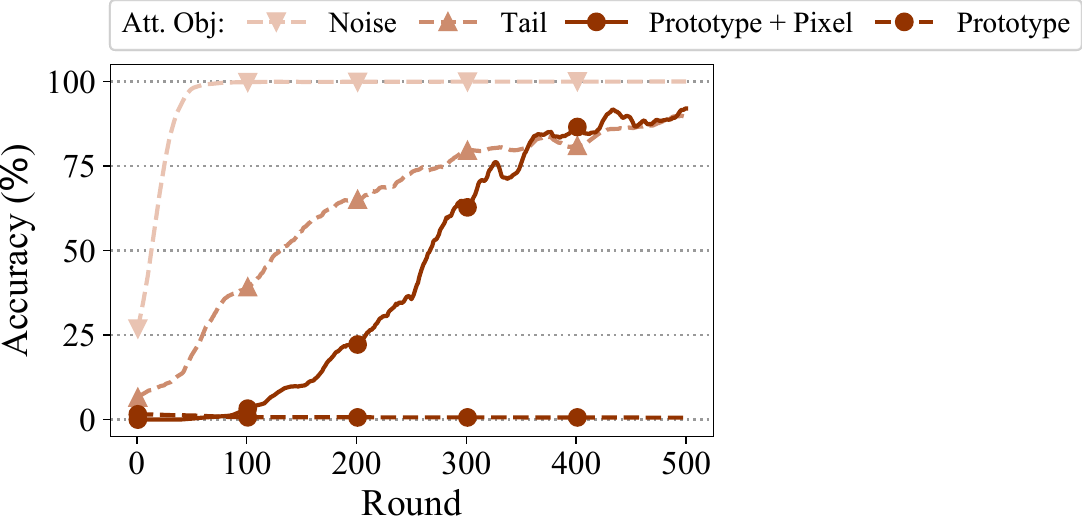}
        \caption{\taskI}
        \lfig{analysis:edge_case_pixel:fmnist}
    \end{subfigure}
    \hfill
    \begin{subfigure}[b]{0.485\columnwidth}
        \vskip 0pt
        \includegraphics[height=2.86cm]{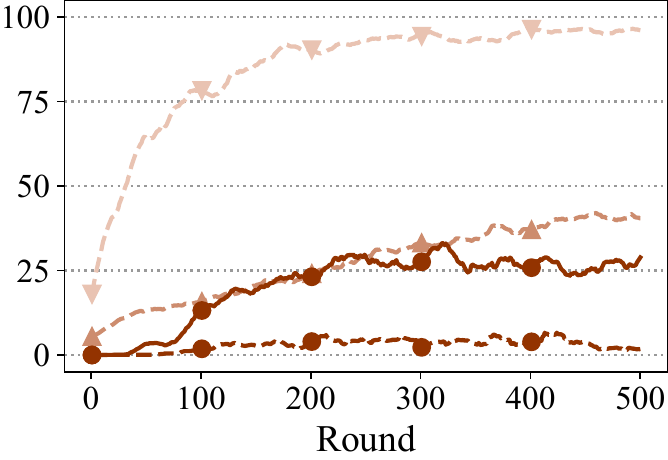}
        \caption{\taskII}
        \lfig{analysis:edge_case_pixel:cifar}
    \end{subfigure}
    \caption{
    	Adding a pixel pattern to prototypical targets makes them behave similar to tail targets.
    \mbox{($\text{M-}r\text{: }1.5$)}
    }
    \vspace{-15pt}
    \lfig{analysis:edge_pixel}
\end{figure}

Norm bounds provide useful robustness guarantees because of the ease of performing practical high-impact single-shot attacks in standard secure \FL.
While it is likely that an attacker will be selected at least once, thus enabling a single-shot attack, being selected consistently over many rounds to enable a continuous attack is unlikely in a practical \FL setup~\cite{shejwalkar2021drawing}.
Achieving this persistence would require an attacker to compromise a significant portion ($>1\%$) of clients, which corresponds to \ourMany of devices
in large-scale \FL deployments.
We thus conclude that norm bounds provide important practical robustness guarantees for \FL.
Although continuous attacks on \ourTailTarget targets remain effective even under norm bounds, they require a significantly stronger attacker, which is less of a concern in most practical deployments.

\subsecspacingtop
\subsection{Discussion}%
\subsecspacingbot
\lsec{discussion}

As ML is deployed in a wider range of settings, it has become clear that it must move beyond simply pursuing accuracy and start to tackle important objectives that matters in real-world deployments such as privacy and robustness.
\FL, though presenting many privacy benefits, amplifies ML robustness issues by exposing the learning to an active attacker that can be present throughout the training process and fine-tune their attacks to the current state of the model.
In consequence, we have seen new and powerful attacks emerge that effectively \mbox{impact \FL integrity.}
In this analysis, we show that norm bounding can provide meaningful practical robustness guarantees for FL.
Furthermore, our analysis shows that many attacks that remain possible in the presence of norm bounds are likely inherent to the learning process.
For example, norm bounds are not able to prevent an attack targeting the tail of the input distribution without sabotaging the model's ability to learn less representative data points.
In general, defenses against all possible attacks are likely an unattainable goal, and we should not refrain from strengthening our systems with existing solutions while we continue to expand our understanding of the underlying robustness issues of decentralized learning. This mirrors trends in the ML community, moving from a binary view of robustness to a more differentiated way of addressing adversarial behavior.

The requirement for ML algorithms to memorize in order to learn \ourTailTarget subpopulations' specific behavior~\cite{Feldman2019-memoization-mf}
has largely been studied in the context of privacy.
However, as we have shown in this analysis, it also has significant implications for ML integrity.
Consequently, mechanisms that have been employed to limit memorization, such as 
appropriate regularization and noise addition (i.e., differential privacy) are also likely to improve 
ML robustness for tail targets~\cite{Feldman2019-memoization-mf}.
However, these countermeasures
introduce various tradeoffs with respect to the accuracy, privacy,
robustness, and fairness{~\cite{bagdasaryan2019differential} and are beyond the scope of this paper.

The appeal of norm bounds is that they exist at an intersection of effectiveness and efficiency where they
prevent an extensive range of simple-to-execute yet devastating attacks in practice while remaining efficiently implementable.
In addition, they do so in a robust way that does not require obfuscation (i.e., the protection still holds if the attacker knows the bound). Norm bounds increase robustness immensely in practical settings, and more extensive protections currently require sacrificing privacy,  efficiency, or both~\cite{byz-ml-krum,byz-ml-trimmed-mean, byz-with-reinforced, byz-defenses-draco,byz-defenses-detox, Shen2016-lf,Fung2018-nl}.
However, the privacy-preserving nature of secure-aggregation-based FL means that enforcing norm bounds is non-trivial.
Relying on clients to self-regulate is not viable in the presence of actively malicious attackers,
and using generic secure computation tools would be prohibitively expensive.
Therefore, we need solutions that can enforce norm bounds over (private) updates from untrusted clients with an overhead small enough to enable practical deployment. The remaining of this paper aims to address this challenge.

\secspacingtop
\section{\oursystem Design} %
\secspacingbot
\lsec{design}

\newcommand{\norm}[1]{\left\lVert#1\right\rVert}

\begin{figure}[t]
    \centering
    \vskip 0pt
    \includegraphics[width=.96\columnwidth]{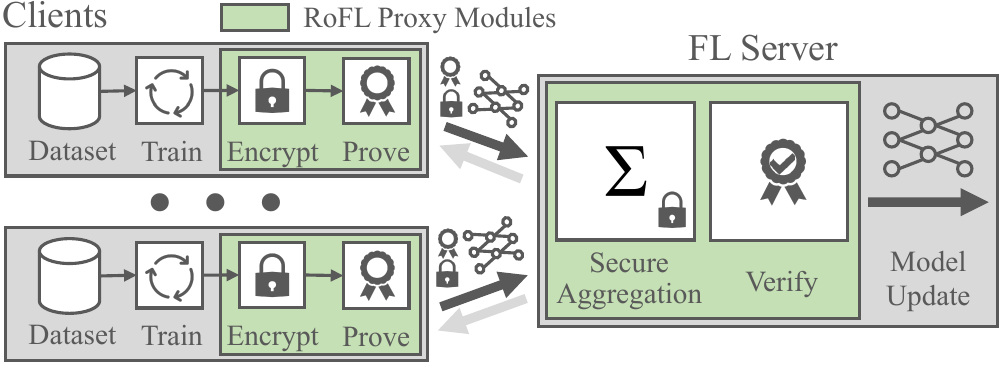}
	\caption{\oursystem augments secure \FL with proof and verification steps to verify constraints on private model updates.}
    \lfig{design:ml:pipeline}
    \vspace{-15pt}
\end{figure}

In this section, we present \oursystem, a new secure aggregation protocol that allows the server to verify that  clients' secret inputs satisfy a predefined constraint (i.e., input validation). We start the section with a brief overview of the cryptographic blocks used in \oursystem, describe our setup and threat model, and then present the details of our protocol.

\subsecspacingtop
\subsection{Preliminaries}
\subsecspacingbot
\lsec{crypto-tools}

\fakeparagraph{Commitments.}
A hiding non-interactive commitment scheme is an algorithm that allows an entity to commit to a value $v$ while keeping it hidden, with the option of later revealing the value.
Given a function $Com : M \times R \rightarrow C$, from a message space $M$ and a randomness space $R$ to a commitment space $C$,
we compute a commitment  to a value $v \in M$ as $c = Com(v, r)$ for $r \gets R$.
Opening a commitment by revealing $r$ and $v$ allows the commitment to be verified by checking $c_v  = Com(v,r)$.
A commitment is \emph{hiding} if it does not reveal any information about the committed value $v$.
Furthermore, a commitment is \emph{binding} if it is (computationally) infeasible to find $v,r$ and $v',r'$ such that $v\neq v'$ but $Com(v,r) = Com(v',r')$.
\fakeparagraph{Zero-Knowledge Arguments of Knowledge.}
A ZKP is a statement that proves a fact to another entity without revealing additional information.
For example, the prover might want to show that they know the discrete log $x$ of a group element $g^x$ and that $x$ lies in a specific range without revealing $x$.
Traditionally, the prover would engage in an interactive challenge-response protocol with the verifier to prove a property about its private data.
Using the Fiat–Shamir transform~\cite{fiat-shamir}, these interactive ZKP protocols can be converted into a non-interactive zero-knowledge proof (NIZK)~\cite{Blum1988-sl}.
The prover generates a single binary string $\pi$ that can be used by any entity to verify the claim without further interaction.
For brevity, we refer to~\cite{Bunz2018-mi} for formal definitions of the security \mbox{properties of NIZKs.} %

\subsecspacingtop
\subsection{Setup and Threat Model}
\subsecspacingbot

\oursystem is designed for a standard FL deployment featuring a single server coordinating a model learning process with a large number of clients (\rfig{design:ml:pipeline}).
In each round of the training process, the server randomly selects $m$ available clients and communicates the current global model to them. 
Each selected client $i$ for $i \in \{1,2, \dotsc,  m\}$ then performs local model training and computes an $\ell$-dimensional update vector $\mathbf{w}_i \in \mathbb{Z}_q^\ell$ that they encode using a homomorphic masking scheme before sending it to the server.
The server uses secure aggregation to combine the clients' update vectors $\sum_{i=1}^m \mathbf{w}_i$ into a single update vector for the global model.
\oursystem allows the server to verify that the client-updates $\mathbf{w}_i$ are valid, i.e., fulfill a predefined constraint while preserving the guarantees of secure aggregation and remaining compatible with the conventional deployment scenario.
Specifically, it allows the $L_p$ norm of client-updates $\mathbf{w}_i$ to be constrained, i.e., $\lVert \mathbf{w}_i \rVert_p < B$ for some $B \ll q$, and we provide concrete instantiations for both $L_\infty$ and $L_2$ norm bounds.
Upon receiving the client input, the server verifies that the proof is valid.
If the verification fails, the server discards the client input, and flags the client as corrupt.

\fakeparagraph{Threat Model.}
We distinguish between a malicious and a semi-honest server.
Malicious parties can deviate from the protocol arbitrarily, whereas a semi-honest party does not deviate from the protocol but analyzes received messages to try and gain information about clients' private inputs.
We inherit the confidentiality guarantees of secure aggregation~\cite{Bonawitz2017-xi,fl-secure-aggr-optimization}, which ensures input privacy for a malicious server that colludes with a fraction $\gamma m$ of malicious clients.
Specifically, an adversary cannot learn anything beyond what they can derive from the aggregation output and the proven constraint of the inputs.
Further \oursystem additionally ensures correctness against malicious clients for a semi-honest server.
We assume a PKI, which allows the server to facilitate end-to-end encrypted and authenticated communication between clients without the need for direct communication between the clients.

\subsection{Secure Aggregation with Commitments}
\subsecspacingbot
\lsec{secure-aggr-comittments}

\fakeparagraph{Secure Aggregation.}
We begin by describing the secure aggregation protocol first introduced by Bonawitz et al.\cite{Bonawitz2017-xi} and extended by Bell et al.~\cite{fl-secure-aggr-optimization}.
After this brief introduction, we show how we extend the protocol to support input constraints.
The secure aggregation protocol relies on a homomorphic masking scheme, 
where each client blinds their input vector $\mathbf{w}_i$ with a key $\mathbf{r}_i$, which are generated so that they cancel out when aggregated.
The distribution of masking keys $\mathbf{r}_i$ between clients is achieved by the \generateMasks sub-protocol.
In order to support dropouts from clients during the aggregation, \reconstructMasks is used to reconstruct mask contributions from clients that dropped out.
\oursystem extends the protocol by adapting the homomorphic masking scheme but does not need to modify \generateMasks and \reconstructMasks.
We, therefore, refer to \cite{fl-secure-aggr-optimization} for a description.

We abstract the method of how clients hide their vectors given a masking key as an encoding scheme in order to improve the presentation of our later extensions.
This encoding scheme (\funEncode, \funDecode) should be homomorphic in both the messages $\mathbf{w}_i$ and the keys $\mathbf{r}_i$ created\footnote{Technically, in \cite{Bonawitz2017-xi}, \generateMasks is used to derive seeds which are then expanded into the keys via a PRG.} } by \generateMasks. Specifically, $\sum_i \funEncode(\mathbf{w}_i, \mathbf{r}_i) = \funEncode(\sum_i \mathbf{w}_i, \sum_i \mathbf{r}_i)$.
The encoding scheme allows the server to aggregate messages encoded with clients' individual keys $\mathbf{r}_i$ and decrypt the aggregate $\funDecode(\sum_i \funEncode(\mathbf{w}_i, \mathbf{r}_i), \mathbf{r})$ iff it has access to the aggregate key $\mathbf{r} = \sum_i \mathbf{r}_i$ which can be computed  securely using \reconstructMasks.
The encoding and decoding functions of the original secure aggregation protocol are defined as:
\begin{equation}
	\begin{split}
		\funEncode(\mathbf{w}_i, \mathbf{r}_i) &\coloneqq \mathbf{w}_i + \mathbf{r}_i \mod q \\
		\funDecode(\mathbf{y}, \mathbf{r}) &\coloneqq \mathbf{y} - \mathbf{r} \mod q
	\end{split}
\end{equation}
for $\mathbf{w}_i, \mathbf{y}, \mathbf{r}_i, \mathbf{r} \in \mathbb{Z}^\ell_q$.\\

\fakeparagraph{Correctness with malicious clients.}
In order to enforce robustness constraints in secure \FL, the encoding scheme needs to support efficient client-side zero-knowledge constraints. 
More importantly, however, the encoding scheme must allow the server to verify \emph{correctness} of the protocol in the face of actively adversarial clients.
The original secure aggregation protocol only guarantees input privacy, but not correctness, which they note ``is much harder to achieve"~\cite{Bonawitz2017-xi}.
For instance, malicious clients can use inconsistent masking values $\mathbf{r}^{\prime}_i$ in \generateMasks and \reconstructMasks that can result in arbitrary outputs at the server.
We must first ensure the correctness of the secure aggregation protocol in this setting if we are to prove any form of robustness constraints meaningfully.

For the server to recover the correct result, encoding keys must add up to the aggregate key $\sum_i \mathbf{r}_i = \mathbf{r}$.
In classical secure aggregation, this would require clients to prove that they executed $\funEncode(\mathbf{w}_i, \mathbf{r_i})$, including the generation of the $\mathbf{r}_i$, honestly as specified by the protocol.
However, proving this is prohibitively expensive, as the formation of $\mathbf{r}_i$ entails key agreements and many evaluations of a PRG.
One of our key insights is that it is sufficient for correctness to ensure that the sum of the keys used in the encodings $\sum_i \mathbf{r}_i$ is equal to the aggregate key output by \reconstructMasks.
However, proofing this condition for the original encoding scheme still introduces considerable overhead.
Instead, we propose an optimization that allows the server to verify this condition efficiently by exploiting the homomorphic nature of the encoding scheme.
We show this optimization for a generic construction before considering how to instantiate it efficiently.

We define a new generic encoding scheme from two homomorphic masking schemes $\E{w}$ and $\E{r}$:
\begin{equation}
	\begin{split}
	&\funEncode(\textbf{w}_i, \mathbf{r}_i) = (\textbf{c}^{(1)}_i, \textbf{c}^{(2)}_i) = (\E{w}(\textbf{w}_i, \mathbf{r}_i),  \E{r}(0, \mathbf{r}_i))\\
	&\funDecode((\textbf{c}^{(1)}_i, \textbf{c}^{(2)}_i), \mathbf{r}) = 
	\D(\textbf{c}^{(1)}_i, \mathbf{r})
\end{split}
\end{equation}
Here, $\E{w}$ and $\E{r}$ must be homomorphic in both message and key, just as the original encoding scheme is. We note that $\E{r}$ must be secure even for a known message, which the original scheme does not fulfill.
Depending on the instantiation, we might also require a proof that the randomness used in  $\textbf{c}^{(1)}_i$ and $\textbf{c}^{(2)}_i$ is the same.
The server can use the second element to verify if the output $\mathbf{r}$ from $\reconstructMasks$ cancels out to zero by comparing $ \E{r}(0, \mathbf{r})$ to $\sum_{i=1}^{m} \mathbf{c}^{(2)}_i$. If they are equal, this implies $\mathbf{r} = \sum_{i=0}^m \mathbf{r}_i$.
Otherwise, the server aborts that particular round and proceeds with another subset of clients.
Hence, \oursystem provides correctness for actively malicious clients by detecting malformed aggregations.

\fakeparagraph{Commitment-based Encoding.}
We identify ElGamal commitments~\cite{elgamal} as an ideal building block in our design because we can use them to instantiate our generic encoding scheme extremely efficiently.
ElGamal commitments are both computationally hiding and information-theoretically binding under a standard discrete log hardness assumption.
Furthermore, ElGamal commitments are additively homomorphic in both the messages and keys, thus supporting the required homomorphisms.
In addition, efficient zero-knowledge range proofs~\cite{Bunz2018-mi} exist that are
a natural fit to realize p-norm constraints and operate directly on (parts of) the ElGamal commitment.

We lift the masking approach from secure aggregation to the commitment-based setting by instantiating our generic encoding scheme with a \emph{single} ElGamal commitments $Com_{EG}(\mathbf{w}_i, \mathbf{r}_i)$ (for each element in $\mathbf{w}_i$)  by setting
\begin{equation*}
	\begin{split}
		(\texttt{E}_\texttt{w}(\mathbf{w}_i, \mathbf{r}_i), \texttt{E}_\texttt{r}(0, \mathbf{r}_i)) &= Com_{EG}(\mathbf{w}_i, \mathbf{r}_i) = (g^{\textbf{w}_i}h^{\mathbf{r}_i}, g^{\mathbf{r}_i}) \\
		\texttt{D}_\texttt{w}(\mathbf{y}, \mathbf{r}) &= \textit{dlog}_g\left(\mathbf{y} \cdot h^\mathbf{-r}\right)
	\end{split}
\end{equation*}
where $\textbf{w}_i, \textbf{w}, \mathbf{r}_i, \textbf{r} \in \mathbb{Z}_q^\ell$ and $g, h$ are generators in a group $\mathbb{G}$ of prime order $q$. Note that $g^{\textbf{w}_i}h^{\mathbf{r}_i}$ is also a valid Pedersen commitment~\cite{Pedersen1992-vh} to $\textbf{w}_i$ in its own, and we will occasionally make use of this fact in the rest of this section.
The decoding function $\texttt{D}_\texttt{w}$ computes the discrete logarithm with respect to $g$, which results in $\textbf{w}$ if $\mathbf{y}$ is a correct encoding of $g^{\textbf{w}}h^{\mathbf{r}}$.
We assume that the discrete log problem is hard in $\mathbb{G}$.
While this means that calculating the discrete logarithm of $y = g^x$ is hard for generic $x \in \mathbb{Z}_q$, 
the aggregation result space in our domain is small (e.g., a 32~bit integer) and hence the discrete logarithm can be computed efficiently~\cite{Shafagh2017-rx,Pollard1978-kk, Shi2011-dw}.
Clients generate a standard non-interactive proof-of-knowledge of discrete logarithm~\cite{camenisch1997proof} to ensure the commitments are well formed, i.e., clients use the same $\mathbf{r}_i$ in $g^{\textbf{w}_i}h^{\mathbf{r}_i}$ and $ g^{\mathbf{r}_i}$.

\fakeparagraph{Alternative approaches.}
We compare the performance of our ElGamal-based encoding with an alternative candidate using RLWE-based encryption~\cite{Brakerski2014-uq}, instantiating our generic encoding as
\begin{equation*}
	\begin{split}
		\E{w}(\textbf{w}_i, \mathbf{r}_i) &\coloneqq (\text{ct}_0, \text{ct}_1) = (\mathbf{a} * \mathbf{r}_i + \mathbf{w}_i + t * \mathbf{e}, -\mathbf{a}) \\
		\E{r}(\mathbf{r}_i) &\coloneqq (\text{ct}'_0, \text{ct}'_1) = (\mathbf{a}' * \mathbf{r}_i + 0  + t * \mathbf{e}', -\mathbf{a}') \\
		\D(\text{ct}_0, \text{ct}_1) &\coloneqq   \left( \text{ct}_0 + \text{ct}_1 * \text{sk} \right) \mod t 
	\end{split}
\end{equation*}
where the masks $\mathbf{a}$, $\mathbf{a}'$ are random ring elements, $\mathbf{e}$, $\mathbf{e}'$ are small noise terms added for security, and $t$ is a public parameter. We refer to \cite{Brakerski2014-uq} for further details and a proof of security.
While this encryption scheme is actually \emph{fully} homomorphic in the message, it is also \emph{additively} homomorphic in the keys.
We implement this using the state-of-the-art Microsoft SEAL library.
\rtab{protocol_comparison} shows that our ElGamal construction outperforms the RLWE-based construction independently of the proof system used to enforce the input constraints.

\begin{table}
    \resizebox{\columnwidth}{!}{%
    \begin{tabular}{lcccc}
        \toprule
        \multicolumn{1}{l}{} & RLWE + & EG + & RLWE + & EG + \\
        \multicolumn{1}{l}{} & Groth16 & Groth16 & BP & BP (\rsec{design}) \\
        \midrule
        \multicolumn{1}{l}{Communication} & $1$ & $1$ & $\log(\ell)$ & $\log(\ell)$ \\
        \cmidrule{0-0}
        \tableindent Vector size & $664$KB & $524$KB & $664$KB & $524$KB \\
        \tableindent Proof size & $0.192$KB & $0.192$KB & $2.01$KB & $1050$KB  \\
        \midrule
        \multicolumn{1}{l}{\multirow{2}{*}{Prover}} & $\ell \log(\ell)$ & $\ell \log(\ell)$ & $\ell$ & $\ell$ \\
        \tableindent & $208.37$s & $610.15$s & $16896.05$s & $5.50$s \\

        \midrule
        \multicolumn{1}{l}{\multirow{2}{*}{Verifier}} & $1$ & $1$ & $\ell$ & $\ell$ \\
        & $0.002$s & $0.01$s & $1140.97$s & $0.57$s \\
        \midrule
        Trusted Setup & $211.71$s & $618.56$s & --- & ---  \\
        \bottomrule
    \end{tabular}
    }
    \caption{
        We implement and evaluate several combinations of our encoding scheme and zero-knowledge proof system to present their concrete communication and computation overhead.
        The concrete rows are based on proving an 8-bit $L_\infty$-norm bound for a vector of $2^{13}$ elements on a c5d.4xlarge AWS instance.
        For the asymptotic rows, \mbox{Big-O} notation is omitted for brevity.
    }
    \ltab{protocol_comparison}
\end{table}

\begin{table}[]
    \centering
    \resizebox{\columnwidth}{!}{%
    \begin{tabular}{ccccccc}
        \toprule
        \multicolumn{3}{l}{} & \multirow{2}{*}{Plaintext} & Bonawitz et al. & Bell et al. & \oursystem \\
        \multicolumn{3}{l}{} && \cite{Bonawitz2017-xi} & \cite{fl-secure-aggr-optimization} & (\rsec{design}) \\
        \midrule
        \multicolumn{3}{l}{Client communication} & $\ell$ & $\ell + m$ & $\ell + \log(m)$ & $\ell + \log(m)$ \\
        \midrule
        \multicolumn{3}{l}{Client computation} & $\ell$ & $\ell m$ & $\ell \log(m)$ & $\ell \log(m)$ \\
        \midrule
        \multicolumn{3}{l}{Input privacy} & \Circle & \CIRCLE & \CIRCLE & \CIRCLE \\
        \multicolumn{3}{l}{Aggregation correctness} & \Circle & \Circle & \Circle & \CIRCLE \\
        \multicolumn{3}{l}{Input validation} & \CIRCLE & \Circle & \Circle & \CIRCLE \\
        \bottomrule
    \end{tabular}
    }
    \caption{Overview of computation and communication asymptotics and security properties for different (secure) aggregation algorithms.
    Big-O notation is omitted for brevity.
    }
\end{table}

\subsecspacingtop
\subsection{ZKP of Norm Bounds}
\subsecspacingbot
\lsec{zkp-norm-bounds}
In order to enforce norm bounds, \oursystem requires each client to provide a NIZK proof that their update vector $\mathbf{c}_i=\funEncode(\mathbf{w}_i, \mathbf{r}_i)$ is well formed and has a norm bounded by $B$, where $B$ is set by the server.
\oursystem offers two methods to select the norm bound adaptively, either the server uses public training data to compute an update and estimate a bound, or the median method presented in \rsec{prototypical-attacks} is followed.
Let $\mathbf{w}_i$ be the parameter vector and $\mathbf{r}_i$ the vector of canceling nonces of client $i$.
In Camenisch and Stadler notation~\cite{camenisch1997proof}, a client needs to provide a proof of the following relation to the server:
$$NIZK(\mathbf{w}_i,\mathbf{r}_i)\{\C = (g^{\mathbf{w}_i} h^{\mathbf{r}_i}, g^{\mathbf{r}_i}) \wedge \lVert \mathbf{w}_i \rVert_p < B\}$$
A wide range of ZKP systems exists to implement such proofs with various performance and security trade-offs.
For \FL workloads with a large number of parameters, we require a proof system that is as lightweight as possible on the client, has moderate bandwidth requirements, and has efficient realizations of range proofs compatible with the homomorphic commitments required for norm constraints.
\oursystem uses Bulletproofs~\cite{Bunz2018-mi} because they provide linear time complexity for prover and verifier, they only have a logarithmic proof size, they do not rely on a trusted setup, they have a specialized efficient realization of range proofs with batching, and, lastly, they operate directly on the first component of the ElGamal commitment.
We compare the performance of \oursystem with Bulletproofs with Groth16~\cite{Groth2016-cu}, a highly efficient state-of-the-art ZKP systems that offers (extremely small) constant proof sizes and verification times. Note that this efficiency comes at the cost of requiring a trusted setup.
We use Circom~\cite{Garcia_Navarro2020-jg} to define a ZKP circuit for each encoding in its custom specification language.
Circom then translates this into a system of rank-1 constraints (R1CS) which both Groth16 and Bulletproofs support.
Groth16 tends to outperform Bulletproofs in terms of concrete prover time for generic R1CS proofs.
However, since \oursystem uses rangeproofs over Pedersen commitments, which can be instantiated much more efficiently in Bulletproofs than generic R1CS proofs, they significantly outperform Groth16 for this task.
However, bridging the gap between the Pedersen commitment that Bulletproofs natively operate on and the ElGamal commitments we use requires a proof of well-formedness that introduces virtually no computational overhead (less than 0.15~s of the prover runtime are due to the well-formedness proof) but significant communication overhead, accounting for all but 1KB of the proof size of our system.
However, the reduction in prover time in combination with very competitive verifier times more than makes up for this in practice as we show in \rsec{eval}.

We now show how to realize $L_\infty$ and $L_2$ constraints; the two variants of norm constraints that \oursystem supports. 
We start with the $L_\infty$ norm bound where we can consider each parameter independently.
We then show how to bound the $L_2$ norm, which depends on all parameters in the vector and which requires clients to additionally commit to, and prove range bounds over, the squares of each parameter.
Although this introduces significant additional costs in communication and computation, $L_2$ bounds are frequently used in the ML robustness literature because the lack of strict bounds on each parameter allows for more uneven parameter weight distributions in the updates.

\fakeparagraph{$\mathbf{L_\infty}$-Norm.}
To bound the $L_\infty$ norm by $B$, it is sufficient for the prover to show to the verifier that each parameter $w_j \in \mathbf{w}_i$ is within a bounded range $w_j \in [0, B)$.
For each committed parameter $c_j \in \C $, the client provides a ZKP that $w_j \in [0,B)$ and that $c_j$ was generated correctly.
\oursystem combines a proof-of-knowledge of the discrete logarithm~\cite{camenisch1997proof} for the correctness of the commitment and a specialized efficient range-proof from Bulletproofs~\cite{Bunz2018-mi} for the range check.
Bulletproofs can aggregate all $\ell$ required range proofs, one for each parameter update, into a single proof %
 consisting of only $2(\log_2(b) + \log_2(\ell)) + 4$ group elements where $b$ is the bit length of the range.
In \FL, this reduces the bandwidth cost from linear in the number of parameters to logarithmic (e.g., an improvement of 26000x for 100k parameters).
Further, instead of mapping 32-bit floating-point parameters directly to $\mathbb{Z}_q$ with standard fixed-point encoding, \oursystem compresses parameters into $b$-bit integers (e.g., $b=8$ or $b=16$) with probabilistic quantization~\cite{fl-update-compression}.
Our evaluation in \rsec{eval} shows that this quantization does not lead to a significant loss of accuracy in \mbox{the overall model.}
\fakeparagraph{$\mathbf{L_2}$-Norm.}
Bounding each parameter by $B$ using the $L_\infty$ bound implies a trivial bound of the $L_2$ norm to $\ell \cdot B^2$.
However, since an $L_2$ norm bound is supposed to allow more flexibility for the individual parameters, this is not a useful way to instantiate $L_2$ bounds.
On the other hand, it is not sufficient to merely proof  $\sum_{i=1}^\ell w_i^2 < B_{L_2}^2$.
Because we are working in $\mathbb{Z}_q$ and a malicious client could cause the sum to wrap around to small values by setting a $w_j$ sufficiently large to cause an overflow.
Therefore, we need to prove that each element is sufficiently small to prevent this attack.
We note that each honest parameter must satisfy $w_j^2 < B_{L_2}^2$ because otherwise the sum would trivially violate the $L_2$ bound.
Therefore, we can enforce an $L_\infty$ bound of $B_{L_2}$ without loss of generality.
We therefore construct our $L_2$ norm bounds as an extension of our $L_\infty$ norm bounds, appending an additional proof to verify that the sum of squared parameters is bounded. 
More specifically, for each commitment $c_j$ to parameter $w_j$, the client provides an additional Pedersen commitment to the square of the parameter, $c^{\prime}_j \coloneqq Com_{PD}(w_j^2, r^{\prime}_j) = g^{{w}_j^2} h^{r^{\prime}_j}$ and provide a proof that the sum of all $c^{\prime}_j$'s in the vector lies in the range $[0, B_{L_2}^2)$.
In addition, the clients also generate a proof of consistency, to ensure that  $c^{\prime}_j$ indeed commits to the square of the value committed to in $c_j$.
This can be done with an additional non-interactive proof-of-knowledge of the discrete logarithm~\cite{camenisch1997proof} by proving the knowledge of an opening to a Pedersen commitment by rewriting it to $c^{\prime}_j = c_j^{w_j} h^{r^{\prime}_j - w_j r_j}$.
The verifier checks this by computing the same sum over all $c^{\prime}_j$ using the homomorphic property and checking the range-proof.
Again, we use Bulletproofs to instantiate the range proof. %

\fakeparagraph{General Constraints.}
In addition to the $L_\infty$ and $L_2$ norm bounds,
\oursystem can support arbitrary client-side constraints expressed as circuits by using standard general-purpose zero-knowledge proofs supported by Bulletproofs.
However, the optimizations proposed in this paper to make the ZKPs feasible for \FL workloads might not be directly compatible with arbitrary constraints, where other tailored optimizations might need to be considered.

\subsecspacingtop
\subsection{Optimizations}
\subsecspacingbot
\lsec{optimizations}
So far, we have described the cryptographic building blocks of \oursystem and how our system can achieve its performance
through a careful co-design of cryptography and \FL protocol. %
Despite these improvements, scaling the protocol to realistic neural network models remains challenging.
Even as we optimize the cryptographic protocol to the specifics of \FL, the sheer number of proofs required due to the number of model parameters
will result in significant overhead in both computation and bandwidth.
Thus, we explore how we can reduce the number of proofs required
by considering optimizations that take advantage of the combination of our protocol with the underlying \FL model.
First, we consider probabilistic range checks to optimize the $L_\infty$ variant of our protocol.
Second, we present the application of a compression technique based on random subspace learning, compatible with our $L_2$ constraint.
Finally, we discuss how we can use optimistic continuation to interleave the server-side verification of the previous round with client-side training for the next round.
This significantly reduces the effective wall-time overhead at the cost of requiring a one-round rollback in case verification fails \mbox{due to malicious activity.}

\fakeparagraph{Probabilistic Range-Checking ($L_\infty$).}
For the $L_\infty$ bound, the range proofs are by far the most significant component of the computation overhead~(\rsec{eval}).
In the base protocol, the client proves that each element in $\mathbf{w}_i$ is smaller than the bound $B$.
However, it is not necessary to check all $\ell$ elements to detect a malicious client with high probability.
\oursystem can employ probabilistic checking of a random subset of the update to reduce the number of range proofs required.
The key insight is that the server can choose the random subset that it wants to verify after the clients have uploaded their commitments.
Because the commitments are binding, the clients are unable to change the already committed parameters.
A single failed verification in the update is sufficient to detect a malicious client, which means that even comparatively small subsets provide strong guarantees.
    The probability that the server selects at least one element in which the malicious update exceeds the bound follows a hypergeometric distribution.
    We show that, due to the properties of the hypergeometric distribution,
    we can make the probability of a violation being missed vanishingly small.%
For example, assume that a fraction $p_v \in (0,1]$ of the $\ell$ parameters in an update exceeds the bound and the server checks a fraction $p_c \in (0,1]$ of the parameters.
The probability that the server does not detect a malicious update with $p_v \cdot \ell$ elements above the bound, while checking $p_c \cdot \ell$ elements,
can be modeled as $Hyp(p_c \cdot \ell | \ell, \ell \cdot (1-p_v), p_c \cdot \ell)$.
The security guarantees of this optimization depend on the ratio of checked parameters to above-bound malicious parameters.
Although one cannot exclude the possibility of a successful attack modifying a very small number of parameters, down to only 1,
such a hypothetical attack would have to be incredibly sophisticated, because the attacker is significantly more constrained than with existing attacks.
As we move to apply this optimization to larger models, the fraction of checks required remains small even as the total number of parameters increases.
We provide an empirical analysis of the security of probabilistic checking under generic and adaptive attacks in Appendix~\rsec{pc}.

Unfortunately, this optimization cannot be applied in the $L_2$ norm variant, because even a single out-of-bound parameter in the update vector could lead to an overflow of the sum-of-squares proof.
Essentially, this would allow an adversary to submit an arbitrary update vector while remaining undetected with high probability.

\fakeparagraph{Compression Techniques ($L_2$).}
Because both computation and communication overheads are linear in the size of the update vectors~(\rsec{eval}), we can benefit from ML~\cite{fl-update-compression} compression techniques that reduce the size of the updates.
However, the need to be compatible with secure aggregation and norm bounding limits the number of applicable techniques.
In \oursystem, we consider random subspace learning~\cite{subspace-ml} as an optimization, which was initially introduced to help understand the hardness of ML tasks.
Random subspace learning applies an $L_2$-norm-preserving transformation that reduces the number of parameters required for a model update.
The transformation preserves the $L_2$-norm, and thus the technique is compatible with our $L_2$ constraint.
In random subspace learning, the model is trained in a random subspace $W^d$ that has a lower dimension $d$ than the original model's dimension $l$.
The random subspace $W^d$ is projected onto the original model space $W^l$ using an orthonormal projection matrix $P \in \mathbb{R}^{l \times d}$:
$W^l = W^l_0 + \mathrm{P}W^d$.
During training, $W^l_0$ and $P$ are treated as constants, and optimizations are done on $W^d$.
Therefore, only $W^d$ has to be exchanged between the client and server in each round, resulting in an improvement in the number of parameters from $l$ to $d$.
The parameter $d$, referred to as the \emph{intrinsic dimension}~\cite{subspace-ml}, defines the compression ratio and has to be set by the server depending on \mbox{the training task.}

\fakeparagraph{Optimistic Continuation.}
The computation time at the server is dominated by proof verification~(\rsec{eval}).
However, proof verification does not necessarily need to block the system from running the next training round because the server already has access to the aggregation result, i.e., the global model for the next round.
The clients' training usually takes significant time, depending on the complexity of the model and the size of the client dataset.
In \oursystem, we leverage this insight to
execute the client training and the server proof verification in parallel.
On receiving all client update vectors, the server optimistically aggregates the vectors, 
attempts to decode
the sum with the aggregate key from \reconstructMasks, and proceeds to the next round before verifying the proofs. 
The server then verifies the proofs of the previous round while the model training of the next round is already underway.
If the verification of the previous round succeeds, the server takes part in \reconstructMasks to reconstruct the aggregate key for the current round.
Should there be any inconsistencies, e.g., a bound violation, the server aborts the new round and resets the model.
However, because inconsistencies should be infrequent in practice, this optimistic approach can reduce overall wall-time significantly.
Note that the server waits before initiating the second phase of secure aggregation because otherwise, a malicious client could compromise the privacy of other clients' training data.
In this scenario, clients could train on a model of which the integrity is not yet verified by the server, allowing a malicious client to replace the global model in aggregation with a malicious one.
This setting could be exploited by applying a range of recent privacy attacks~\cite{Fowl2021-sn, gradient-privacy-2, Pasquini2021-cc} that are able to extract training data from maliciously altered models.
Delaying \reconstructMasks means that the server cannot decrypt the global update and thus does not gain any additional information compared to the protocol without optimistic continuation.

\vspace{-3pt}
\secspacingtop
\section{Evaluation}
\secspacingbot
\lsec{eval}
    \vspace{-3pt}

\renewcommand{\arraystretch}{1}
\begin{table}[t!]
    \small
    \centering
    \resizebox{\columnwidth}{!}{%
        \begin{tabular}{l l l l l}
            \hline
            & \evalMNIST & \evalCIFARS & \evalCIFARL & \evalShakespeare \\
            \toprule
            \textbf{Network} & CNN & LeNet5~\cite{lenet5} & ResNet-20~\cite{resnet1} & LSTM~\cite{lstm} \\
            \textbf{Weights} & 19k & 62k & 273k & 818k \\
            \textbf{Dataset} & F-MNIST~\cite{federated-mnist} & CIFAR-10~\cite{cifar10} & CIFAR-10~\cite{cifar10} & Shakespeare~\cite{federated-mnist} \\
            \textbf{Task} & image class. & image class. & image class. & text gen. \\
            \hline
        \end{tabular}}
    \caption{Training tasks used in our evaluation.}
    \ltab{tab:evaluation_tasks}
    \vspace{-16pt}
\end{table}
\renewcommand{\arraystretch}{1}

In this section, we quantify the overhead of \oursystem and show that it can be used to train practical ML models.
\fakeparagraph{Implementation.}
We developed an end-to-end prototype of \oursystem that we make available online. %
We implement the framework in Rust and interface it with Python to train neural networks with TensorFlow~\cite{tensorflow}.
For client-server communication, we rely on the Tonic RPC framework~\cite{tonic} with protobuf.
We use the elliptic curve Curve25519 (i.e., 126-bit security) implementation from the \textit{dalek curve25519} library~\cite{daklek-curve-lib} for cryptographic operations.
The library supports \textit{avx2} and \textit{avx512} hardware instructions for increased performance on supported platforms.
The range proofs are implemented with the Bulletproof library~\cite{daklek-bulletproof-lib}, which builds on the same \mbox{elliptic curve library}.

\begin{figure*}[t]
    \begin{subfigure}[b]{0.23\textwidth}
        \vskip 0pt
        \includegraphics[height=2.5cm]{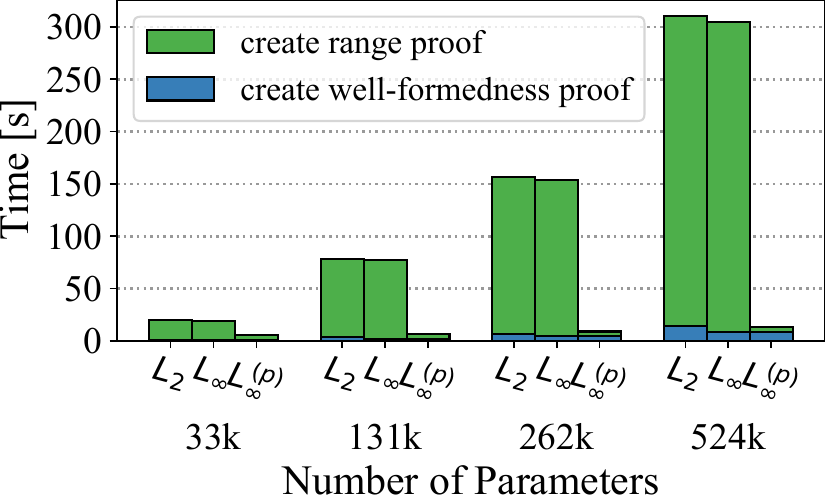}
        \caption{large client}
        \lfig{eval:mbench:clientlarge:zkp}
    \end{subfigure}
    \begin{subfigure}[b]{0.23\textwidth}
        \vskip 0pt
        \includegraphics[height=2.5cm]{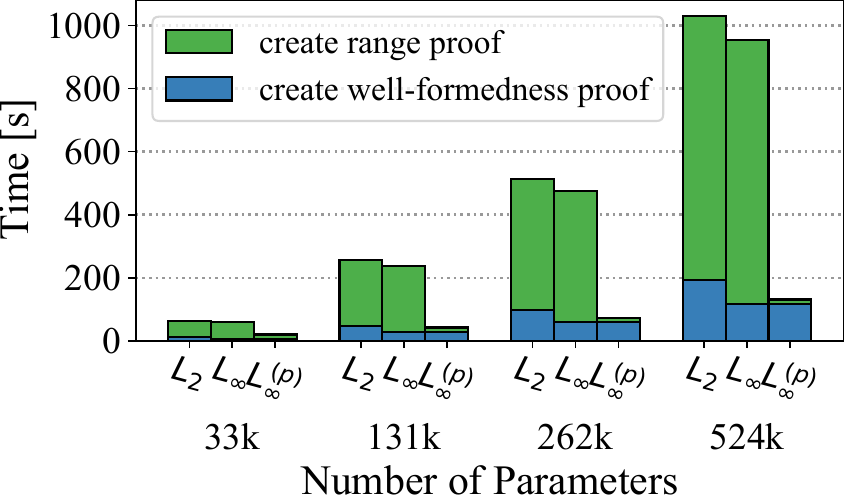}
        \caption{small client}
        \lfig{eval:mbench:clientsmall:zkp}
    \end{subfigure}
    \begin{subfigure}[b]{0.23\textwidth}
        \vskip 0pt
        \includegraphics[height=2.5cm]{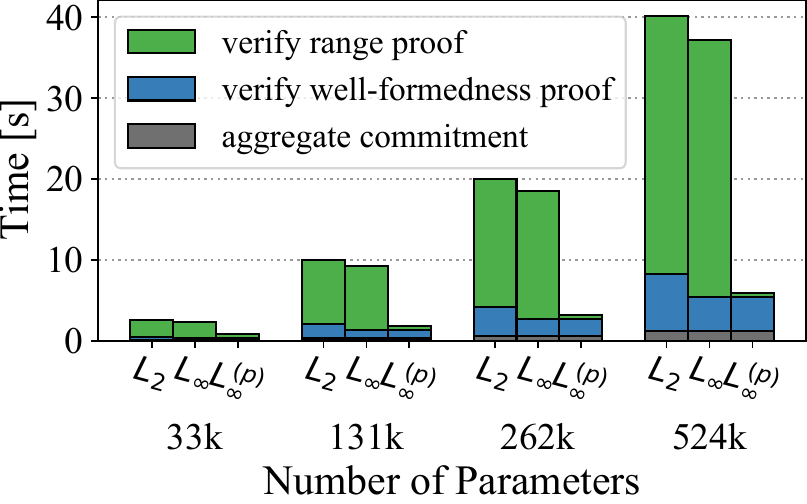}
        \caption{server: per client}
        \lfig{eval:mbench:server:zkp}
    \end{subfigure}
    \begin{subfigure}[b]{0.3\textwidth}

        \begin{subfigure}[t]{0.9\textwidth}
            \vskip -18pt
            \includegraphics[height=0.6cm]{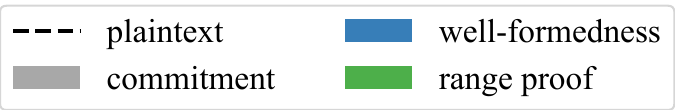}
        \end{subfigure}

        \begin{subfigure}[b]{0.48\textwidth}
            \vskip 0pt
            \includegraphics[height=1.9cm]{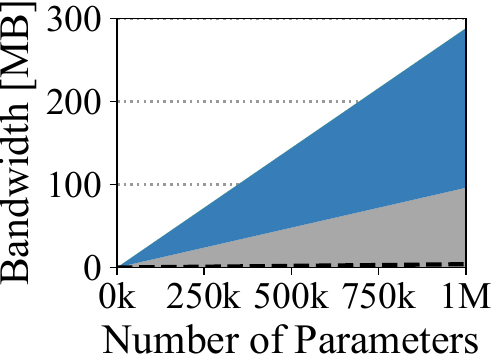}
            \caption{$L_2$ BW}
            \lfig{eval::mbench:bandwidth:l2}
        \end{subfigure}
        \begin{subfigure}[b]{0.48\textwidth}
            \vskip 0pt
            \includegraphics[height=1.9cm]{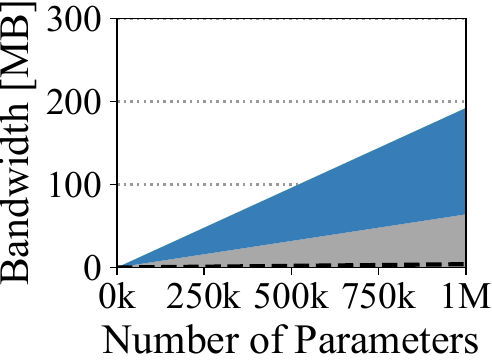}
            \caption{$L_{\infty}$ / ${(p)}$ BW}
            \lfig{eval::mbench:bandwidth:l8}
        \end{subfigure}
    \end{subfigure}
    \caption{Microbenchmark results.
    Client computation cost for creating the $L_2$/$L_\infty$-norm bound proof (a, b).
    Server computation for ZKP verification, commitment aggregation (c).
    Additional bandwidth cost per client to upload the update to the server (d, e).
    }
    \lfig{eval:mbench}
\end{figure*}

\begin{figure*}[t]
    \centering
        \vspace{-9pt}
    \begin{subfigure}[t]{.28\textwidth}
        \includegraphics[width=\columnwidth]{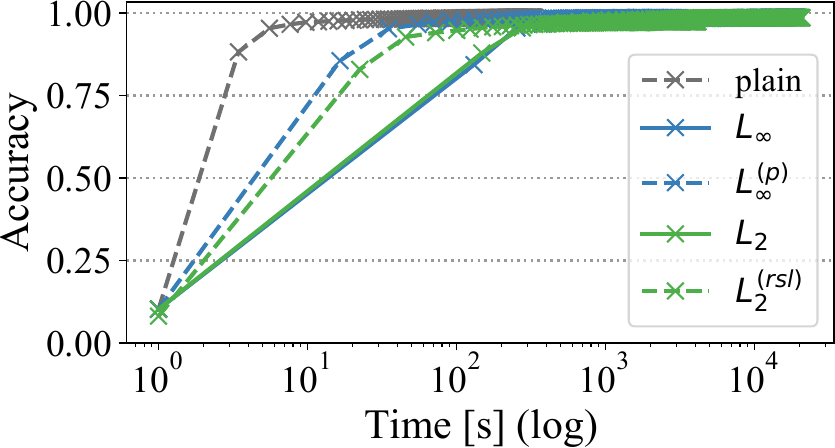}
        \caption{MNIST}
        \lfig{eval::e2e:mnist}
    \end{subfigure} \hfill
    \begin{subfigure}[t]{.28\textwidth}
        \includegraphics[width=\columnwidth]{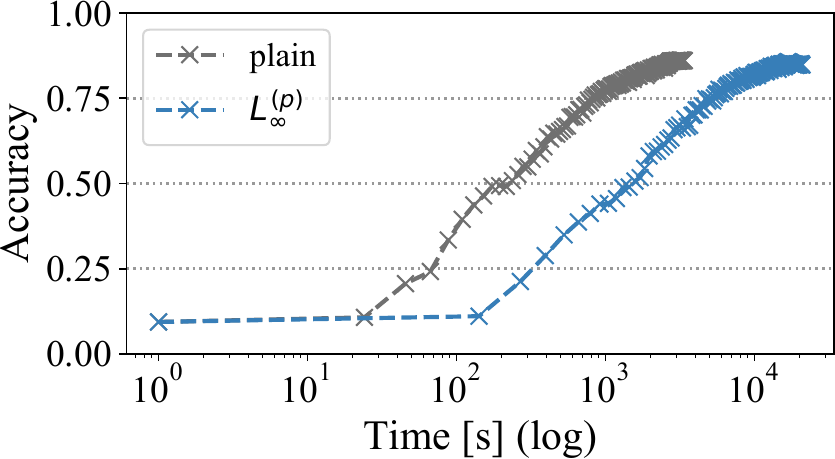}
        \caption{CIFAR-10 L}
        \lfig{eval::e2e:cifar}
    \end{subfigure} \hfill
    \begin{subfigure}[t]{.28\textwidth}
        \includegraphics[width=\columnwidth]{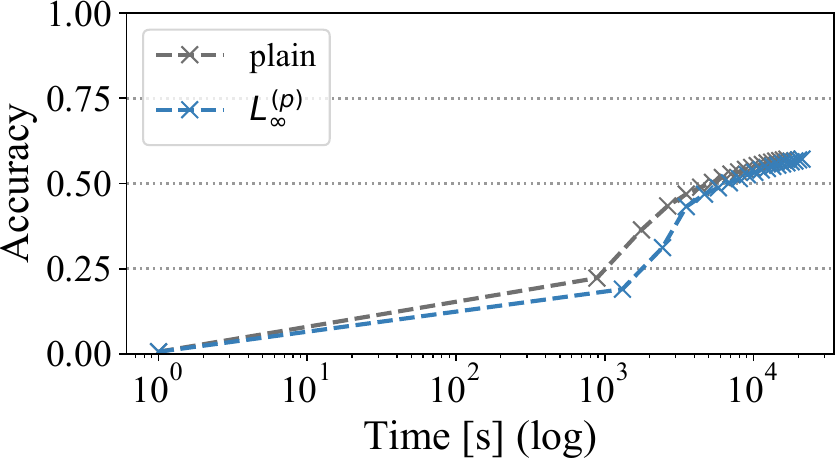}
        \caption{Shakespeare}
        \lfig{eval::e2e:sp}
    \end{subfigure}
    \caption{Model convergence over time.}
    \vspace{-15pt}
    \lfig{eval::e2e}
\end{figure*}

\fakeparagraph{Microbenchmarks.}
We use a single AWS EC2 instance (c5d.4xlarge, 16 vCPU, 32GiB, Ubuntu18.04) with support for \textit{avx512} instructions.
To evaluate the performance of the microbenchmarks on constrained client devices,
we also perform the experiments on a less powerful AWS EC2 instance (t2.medium, 2 vCPU, 2GiB, Ubuntu18.04).

\fakeparagraph{End-to-End benchmarks.}
We evaluate end-to-end system performance for four tasks on the image- and text-based datasets. \rtab{tab:evaluation_tasks} provides an overview of the tasks and models we use in our evaluation.
To simulate a practical deployment, we use five AWS EC2 instances (c5d.9xlarge, 36 vCPU, 72 GiB, Ubuntu18.04) with support for \textit{avx512} instructions and with a latency of 0.5ms between them.
The server uses one instance, and 48 clients are evenly distributed over \mbox{the other four instances.}

\subsecspacingtop
\subsection{Input Validation with ZKP}
\subsecspacingbot
We begin our evaluation by investigating the overhead of integrating ZKPs for the $L_\infty$- or $L_2$-norm bound checks into secure aggregation on clients and the server.
In the following, we first analyze the computation and bandwidth costs of the baseline protocol and then show the benefits of using our probabilistic checking optimization.
We defer discussion of the subspace learning optimization for $L_2$ to the \mbox{end-to-end experiments.}

\fakeparagraph{Computation.}
The client-side computational costs, beyond the \FL protocol itself, only include computing commitments to each parameter and creating the corresponding zero-knowledge norm proofs.
This additional cost is linear in the number of parameters, as seen in \rfigs{eval:mbench:clientlarge:zkp}{eval:mbench:clientsmall:zkp}.
On a strong client (\rfig{eval:mbench:clientlarge:zkp}), the creation of $L_2$ and $L_\infty$ proofs for 262k parameters takes 157 and 154 seconds, respectively, whereas on a less powerful machine (\rfig{eval:mbench:clientsmall:zkp}), this cost increases by 3x to 514 and 477 seconds.
These costs are predominately determined by the range-proof creation, to which the additional well-formedness proofs to ensure correct encoding do not contribute significantly.
Although these costs are significant, they are within an order of magnitude of the model training cost, so clients that can perform the underlying \FL computation efficiently should be able to handle the overhead too.

When using $L_2$-norm proofs, clients need to compute additional proof-of-squares,
which adds a minor overhead compared to $L_\infty$-norm proofs.
Meanwhile, the server side verifies the proofs of each client, aggregates them into the global vector, and reconstructs the aggregated vector once all updates are included.
Similar to the client-side costs, the server-side verification and aggregation costs increase linearly with the number of parameters, as shown in \rfig{eval:mbench:server:zkp}.
With an update size of 262k, the server requires 19--20s per client to verify the proofs and aggregate the commitments, where the verification of the range proofs dominates the processing time.
Once the clients' commitments are aggregated, the server has to solve the discrete-log for each parameter.
The discrete-log solving time increases linearly with the number of parameters (i.e., 390ms for 262k and 464ms for 524k params), but the overhead is marginal compared to proof verification costs.
The tasks at the server are highly parallelizable, so we can reduce the overall computation time by horizontal scaling by adding more processors.

\fakeparagraph{Communication.}
In \rfigs{eval::mbench:bandwidth:l2}{eval::mbench:bandwidth:l8}, we show that %
the bandwidth requirements grow linearly in the number of parameters.
The bandwidth costs are higher for the $L_2$-norm (\rfig{eval::mbench:bandwidth:l2}) than the $L_\infty$-norm (\rfig{eval::mbench:bandwidth:l8}),
due to the additional commitments to the square and the corresponding proofs.
The range proofs can be aggregated into a single proof, whose size increases logarithmically in the number of parameters.
Thus, the bandwidth costs are dominated by the commitment size and the additional group elements required for the well-formedness proof.
For 262k parameters, the data transmitted to the server per client is 75MB (72x compared to plaintext) for the $L_2$-norm setting and 51MB (48x) for the $L_\infty$-norm setting.
Clearly, this is a significant overhead that limits the size of the model that this technique can scale to.
However, it is worth noting that while the relative size increase is dramatic, the absolute size of the updates is
about the same as a few minutes of compressed HD video and, therefore, still well within the capabilities of most modern connections, including wireless and mobile connections.

\fakeparagraph{Probabilistic Checking.}
We now show the performance improvements of probabilistic checking of $L_\infty$ constraints.
We set a maximum failure probability of $10^{-8}$ and assume that at least $p_v = 0.005$ of the parameters in a malicious update must exceed the bound for an attack to be successful. %
We provide a more thorough justification, including empirical evidence, for these choices in Appendix~\rsec{pc}.
Bandwidth is negligibly affected by probabilistic checking because the aggregated range-proofs are already logarithmically sized (\rfig{eval::mbench:bandwidth:l8}).
However, probabilistic checking can reduce the computation cost by 6x--17x for the clients and 6x for the server (\rfig{eval:mbench}).
The optimization shows better improvements when using larger numbers of parameters: for instance, from 154s to 9s on the large client and 19s to 3s per client server-side for 262k parameters.
We conclude that probabilistic checking dramatically reduces the overhead of applying $L_\infty$-norm constraints while offering almost the same \mbox{level of protection.}
\fakeparagraph{Additional Setup Costs.}
These experiments exclude the setup cost for establishing the canceling nonces with Diffie-Hellman key exchanges between the clients.
Compared to the overheads of commitments and ZKPs, these costs are small (i.e., 9KB bandwidth and 24ms computation time per client and round for a 100-client setup).

\subsecspacingtop
\subsection{\oursystem~End-to-End Performance}
\subsecspacingbot
We demonstrate the effectiveness of \oursystem in a practical deployment by showing that \oursystem provides a significant speedup in computation time compared to the unoptimized $L_\infty$- and $L_2$-norm constraints.
For the optimized version of the $L_\infty$-norm, we apply probabilistic checking (denoted by $L^{(p)}_\infty$}, and for $L_2$, we compress the updates with random subspace learning (rsl, denoted by $L^{(rsl)}_2$)~\cite{subspace-ml}.
We also show the feasibility of deploying \oursystem for practical models with a large number of parameters. %

\fakeparagraph{\oursystem $L_2$ vs Secure Aggregation.}
We set the effective number of parameters in $L_2$-norm with rsl, i.e., the intrinsic dimension, for \evalMNIST to 5000 (3.8x parameter reduction), for \evalCIFARS to 12000 (5.2x parameter reduction), and for \evalCIFARL to 40000 (6.8x parameter reduction) as explored by Li et al.~\cite{subspace-ml}.
\rtab{tab:e2e} compares the results of the different runs.
In the \evalMNIST task, our optimization reduces the training time per round from 63x to 13x compared to \evalbaseline.
In the \evalCIFARS and \evalCIFARL experiments, we find that \oursystem reduces the computation per round from 153x to 29x and from 106x to 10x.
Quantization slightly reduces the accuracy for \evalCIFARS and \evalCIFARL,
from $0.61$ for plaintext to $0.60$ for \evalCIFARS, and from $0.86$ to $0.85$ for \evalCIFARL.
In addition, we observe a small impact on accuracy in all tasks due to the subspace ML compression,
which is consistent with the work by Li et al.~\cite{subspace-ml}.
Moreover, the compression reduces bandwidth costs from 37x to 7x for \evalCIFARS and from 36x to 5x for \evalCIFARL compared to the \evalbaseline baseline.

\renewcommand{\arraystretch}{1.15}
\begin{table}
    \centering
    \resizebox{\columnwidth}{!}{%
        \begin{tabular}{lrcrrrcrr}
        \toprule
             &                && \multicolumn{3}{c}{Computation Time} & & \multicolumn{2}{c}{Bandwidth} \\
             \cline{4-6}\cline{8-9}
        Type & Acc. && Round [s] & Total [m] & Factor & & Total [GB] & Factor \\
        \toprule
        \multicolumn{9}{c}{\textsc{MNIST} (19k params, rsl 5k params, 160 rounds)} \\
        SA & 0.99 & & 2 & 6 & 1x & & 1.4 & 1x \\
        $L_2$ & 0.99 & & 131 & 349 & 63x & & 43.0 & 36x \\
        $L_2^{(rsl)}$ & 0.97 & & 26 & 70 & 13x & & 11.2 & 10x \\
        $L_\infty$ & 0.99 & & 122 & 325 & 58x & & 28.9 & 24x \\
        $L_\infty^{(p)}$ & 0.99 & & 20 & 53 & 9x & & 28.9 & 24x \\
        \toprule
        \multicolumn{9}{c}{\textsc{CIFAR-10 S} (62k params, rsl 12k params, 100 rounds)} \\
        SA & 0.61 & & 2 & 3 & 1x & & 2.8 & 1x \\
        $L_2$ & 0.60 & & 280 & 467 & 153x & & 86.9 & 37x \\
        $L_2^{(rsl)}$ & 0.58 & & 53 & 89 & 29x & & 16.8 & 7x \\
        $L_\infty$ & 0.60 & & 251 & 419 & 137x & & 58.3 & 25x \\
        $L_\infty^{(p)}$ & 0.60 & & 38 & 63 & 21x & & 58.3 & 25x \\
        \toprule
        \multicolumn{9}{c}{\textsc{CIFAR-10 L} (273k params, 160 rounds)} \\
        SA & 0.86 & & 21 & 57 & 1x & & 19.7 & 1x \\
        $L_2$ * & 0.85 & & 2250 & 5999 & 106x & & 612.4 & 36x \\
        $L_2^{(rsl)}$ & 0.82 & & 208 & 554 & 10x & & 89.7 & 5x \\
        $L_\infty$ * & 0.85 & & 2216 & 5911 & 104x & & 411.1 & 24x \\
        $L_\infty^{(p)}$ & 0.85 & & 131 & 348 & 6x & & 411.1 & 24x \\
        \toprule
        \multicolumn{9}{c}{\textsc{Shakespeare} (818k params, 20 rounds)} \\
        SA & 0.57 & & 881 & 294 & 1x & & 3.7 & 1x \\
        $L_2$ * & 0.57 & & 5623 & 1874 & 6x & & 114.6 & 37x \\
        $L_\infty$ * & 0.57 & & 5386 & 1795 & 6x & & 77.0 & 25x \\
        $L_\infty^{(p)}$ & 0.57 & & 1114 & 371 & 1.3x & & 77.0 & 25x \\
        \bottomrule
    \end{tabular}
    }
\caption{E2E model training with \oursystem \mbox{(* extrapolated).}
Bandwidth is reported per client per round.}
\vspace{-10pt}
\ltab{tab:e2e}
\end{table}

\renewcommand{\arraystretch}{1}

\fakeparagraph{\oursystem $L_\infty$ vs Secure Aggregation.}
We show the use of probabilistic checking to optimize the $L_\infty$-norm constraint for various model sizes.
Similar to the microbenchmarks, we set a maximum failure probability of $10^{-8}$ in the optimized experiments while assuming $p_v = 0.005$
The optimization reduces the computation overhead in \oursystem per round from 58x to 9x for \evalMNIST, and from 137x to 21x for \evalCIFARS (\rtab{tab:e2e}).
The probabilistic checking optimization shows benefits even for smaller models,
but it shows the most significant speedups when applied to larger models (e.g., \evalCIFARL and \evalShakespeare).
Although bandwidth use remains similar to unoptimized \oursystem,
probabilistic checking reduces the computation overhead for \evalCIFARL by more than one order of magnitude:
from 105x to 6x compared to \evalbaseline.
Meanwhile, for \evalShakespeare, it reduces the computation overhead from 6x to 1.3x.
Because this model has a higher ratio of training computation to the number of parameters,
the cryptographic overhead of \oursystem is smaller.
The larger number of parameters allows probabilistic checking to achieve an even greater relative speedup here.
Model convergence is similar for the \evalbaseline and, as shown in ~\rfig{eval::e2e},
and accuracy reductions due to quantization are less than $1\%$.
As a result, although \oursystem introduces a noticeable overhead, especially in terms of bandwidth, computation and communication in \oursystem remain efficient enough to be practical for real-world deployments.

\fakeparagraph{Optimistic Continuation.}
\rfig{eval:e2e:timings} shows the effect of optimistic continuation.
The server utilizes the idle time by verifying proofs lazily after starting the next round.
The optimization results in a speedup of $39\%$ for \evalCIFARL and $18\%$ for \evalShakespeare by parallelizing proof verification with model training and proof creation.
For the \evalShakespeare, proof verification requires no additional wall-clock time,
because model training and proof creation in $t+1$ take longer than the proof verification of $t$ round.

\begin{figure}[t]
    \centering
    \vskip 0pt
    \includegraphics[width=\columnwidth]{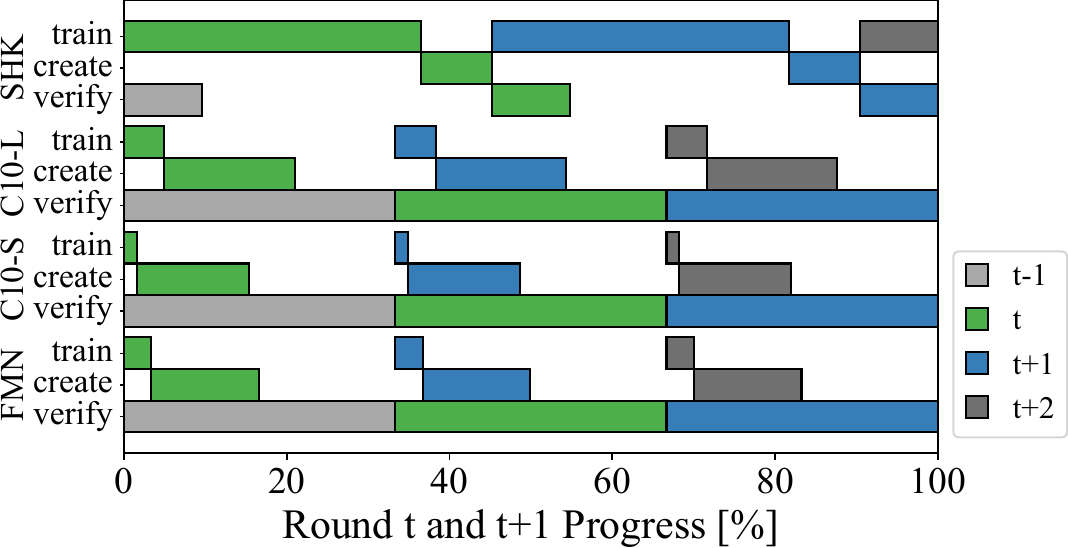}
    \caption{Optimistic continuation allows the parallelization of client-side model training and proof creation with server-side proof verification.}
    \lfig{eval:e2e:timings}
    \vspace{-15pt}
\end{figure}

\secspacingtop
\section{Related Work}
\secspacingbot

In the interest of space, we review related work solely on cryptographic systems similar to RoFL; we have discussed the wider field of related work on \FL robustness in \rsec{sec:analysis}.

\fakeparagraph{Secure Data Collection.}  
Secure aggregation systems~\cite{canny-norm-bound-protocol, secure-aggr-treshold, Castelluccia2009-dr,melis-secure-aggr,dp-with-secure-aggregation,Corrigan-Gibbs2017-kg,Kursawe2011-xy,Popa2011-ko,Shi2011-dw,Acs2011-gg,timecrypt, zeph} enable users to compute sums over their private data without disclosing the inputs. %
A variety of secure aggregation protocols have emerged to meet the needs of a wide range of
 applications that introduce new security, scalability, and performance requirements.
Bonawitz et al.~\cite{Bonawitz2017-xi} and Bell et al.~\cite{fl-secure-aggr-optimization} introduce a secure aggregation protocol designed for FL based on canceling random masks; their protocol builds on~\cite{Acs2011-gg, melis-secure-aggr}.
However, none of these systems address integrity.
The protocol used by \oursystem without ZKP is similar to that proposed by Shi et al.~\cite{Shi2011-dw}; however, ours is tailored to \FL and does not assume a trusted setup. 
Kursawe et al.~\cite{Kursawe2011-xy} extend a variation of the protocol by Shi et al.~\cite{Shi2011-dw} with ZKP to provide correct execution guarantees but does not constrain client inputs. %
Canny et al.~\cite{canny-norm-bound-protocol} propose a private peer-to-peer machine learning platform that allows clients to verify the $L_2$-norm of inputs.
However, it exhibits high overheads that make it unsuitable for the \FL setting. %
A different body of work in this space~\cite{Corrigan-Gibbs2017-kg, Duan2010-kv, prio-all, prio-plus, henry-hh} considers a non-colluding trust model in which the clients split the input values into shares and submit them to non-colluding servers.
This enables efficient input checks, as expensive cryptographic operations can be avoided on the assumption that the servers do not collude.
However, we know from our discussions with \FL industry players that organizations have struggled  %
to deploy this trust model in practice, even when, e.g., using an external non-profit organization.
Thus, the realistic deployment of the non-colluding trust model remains an issue hindering the deployment of these protocols in practice.

\fakeparagraph{FL with Differential Privacy.}
\oursystem opts for a cryptographic solution to hide client updates from the server.
An alternative approach to client update privacy is to employ differentially-private mechanisms that perturb the local client updates~\cite{fl-local-dp1, fl-local-dp2, fl-local-dp3, fl-central-dp1}.
However, a pure perturbation-based approach for privacy in the FL setting that guarantees local differential privacy (i.e., hiding updates) requires clients to add a substantial amount of noise~\cite{fl-local-dp1, fl-local-dp2, fl-local-dp3},
 significantly reducing utility, making it impractical in many situations.
RoFL composes well with differential-privacy techniques for output privacy by enforcing norm bounds on clients' updates
 but one must navigate the trade-off between utility, privacy, fairness, and robustness (\rsec{discussion}).
Still, we believe that perturbation-based differential privacy can be an orthogonal feature that RoFL can be combined with to increase robustness~\cite{dp-robust,Bagdasaryan2018-yx,clipping-sun2019can,edge-case-backdoor} and privacy further.

\privatePublic{}{
\section*{Acknowledgments}
We would like to thank Christian Knabenhans for his help in evaluating RoFL.
We also thank Marko Mihajlovic, Matthias Lei, Emanuel Opel, Kenny Paterson, and the PPS Lab team for their insightful input and feedback.
We would also like to acknowledge our sponsors for their generous support, including Meta, Google, SNSF through an Ambizione Grant No. 186050, and the Semiconductor Research Corporation.
}

\bibliographystyle{IEEEtranS}
\privatePublic{
	\scriptsize{
		\bibliography{biblio}
	}

	\normalsize{
		\appendices

\section{}
\lsec{apx:ssec:additionalexperiments}
We present additional experiments for the robustness analysis in this appendix.

\fakeparagraph{Untargeted attacks.}
\lsec{apx:ssec:untargeted_attacks}
In addition to targeted attacks, we consider the performance untargeted attacks for various norm bounds.
We use the state-of-the-art Projected Gradient Ascent (PGA) attack proposed by Shejwalkar et al.~\cite{shejwalkar2021drawing} in our setup.
The authors show the performance of this attack for different fractions of compromised clients.
We take an orthogonal approach and study the impact of the choice of norm bound on the attack.
We use the same setup as for targeted attacks which allows the attacker to perform an attack for 500 rounds.
We define \emph{Attack Impact} as the difference between the accuracy of the unattacked global model and the attacked model after 500 rounds.
To account for per-round variance of model performance, we compute the accuracy as the average over the last 20 rounds.
For tight norm bounds (\taskI: $L_2$-B $<10$, $r$-M $<15$, $L_\infty$-B $<0.1$, \taskII: $L_2$-B $<20$, $r$-M $<15$, $L_\infty$-B $<0.05$), untargeted attack impact is negligible (i.e., less than $6\%$, \rfig{apx:analysis:untargeted}).
As the bound loosens, attack impact becomes noticeable.
However, when comparing to targeted attacks, untargeted attacks are only effective under a much larger bound than targeted attacks.
For instance, untargeted attacks start to become effective for a median-based bound $r$-M $>=30$, whereas targeted attacks already have significant success for $r$-M $>=15$ (\rfigs{analysis:median:fmnist}{analysis:median:cifar}).
This suggests that strong untargeted attacks are easier to prevent with a norm bound than targeted attacks.

\begin{figure}[t]
    \centering
    \vskip 0pt
    \includegraphics[width=\columnwidth]{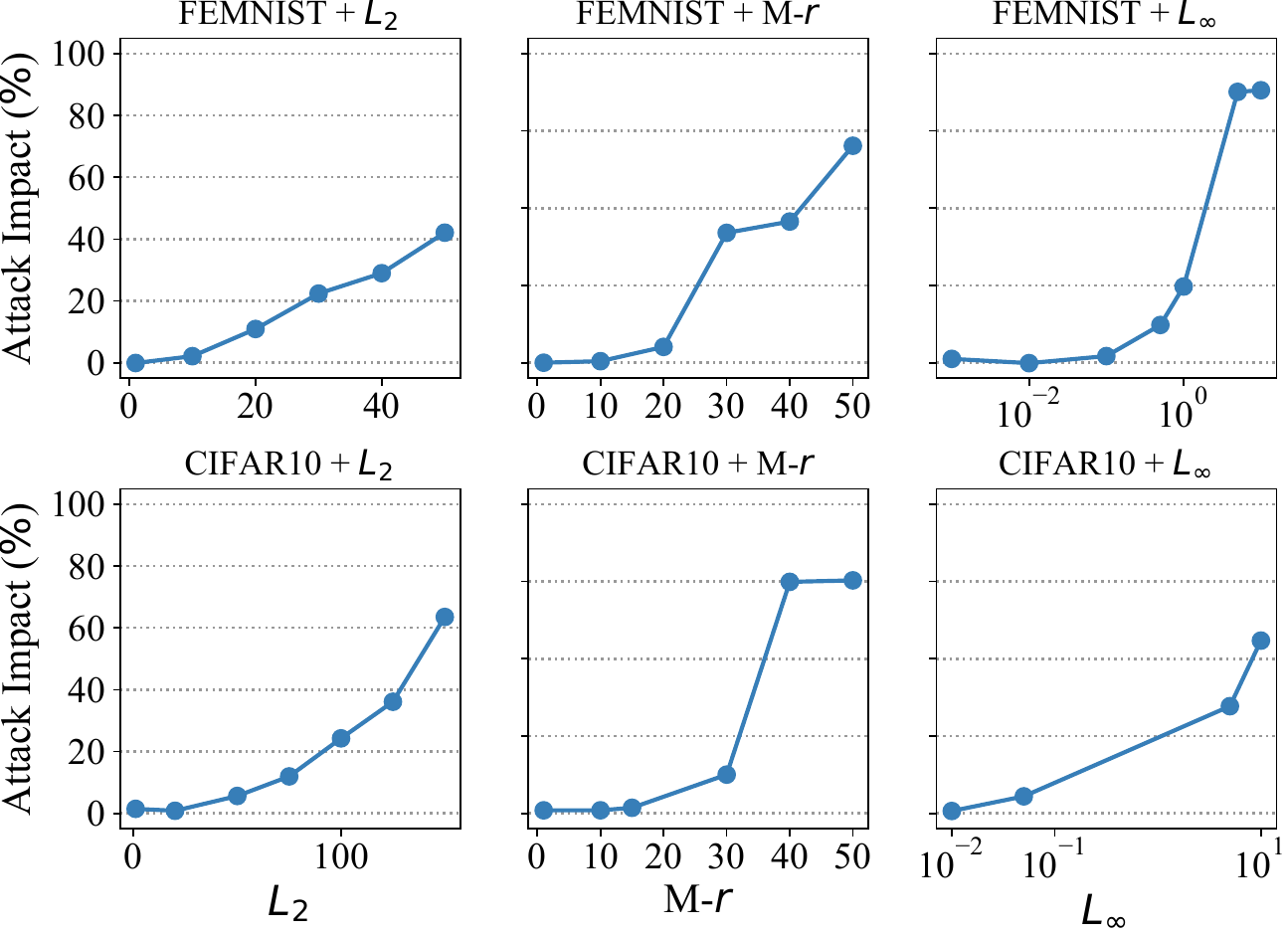}
    \caption{Attack impact of the state-of-the-art untargeted PGA attack.}
    \lfig{apx:analysis:untargeted}
\end{figure}

\fakeparagraph{Comparison of model poisoning attacks.}
We evaluate the three model poisoning attack strategies (\attackAT, \attackPGD, \attackNT) under adaptive norm bounds
    in~\rfig{analysis:compare_attacks:median:all}
    to complement the comparison under static bounds (\rfig{analysis:compare_attacks} in the analysis).
None of the attacks are successful under an appropriate multiplier of the median norm M-$r$ $=1.5$.
    For a looser bound (M-$r$: $15$), the attack strategies are successful and their relative performance diverges,
    with \attackPGD achieving highest malicious accuracy for \taskI and \attackNT for \taskII.

\begin{figure}[t]
    \begin{subfigure}[b]{0.49\columnwidth}
        \vskip 0pt
        \includegraphics[height=3.25cm]{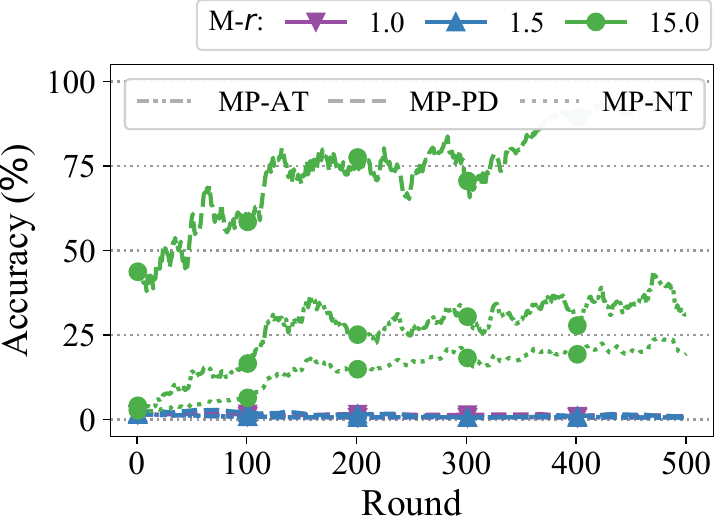}
        \caption{\taskIprototype}
        \lfig{analysis:compare_attacks:median:combined:fmnist}
    \end{subfigure}
    \begin{subfigure}[b]{0.49\columnwidth}
        \vskip 0pt
        \includegraphics[height=3.25cm]{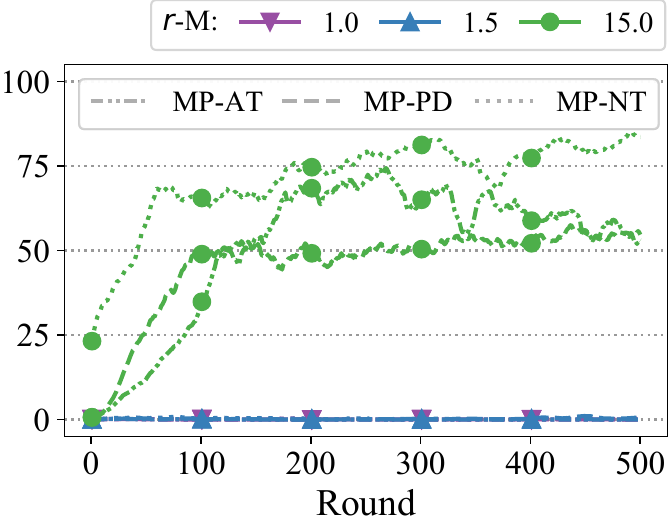}
        \caption{\taskIIprototype}
        \lfig{analysis:compare_attacks:median:combined:cifar}
    \end{subfigure}

    \caption{Comparison of model poisoning attacks under various median bounds.}
    \lfig{analysis:compare_attacks:median:all}
\end{figure}

\fakeparagraph{Growing number of attackers.}
We compare the performance of data poisoning with all three model poisoning in~\rfig{analysis:increase_attackers:all}.
All attacks show similar performance relative to the \% of compromised clients per round,
with \attackAT achieving the highest malicious accuracy improvement over data poisoning (at most 40\%) as shown in Section~\rsec{sec:analysis}.

\begin{figure}[t]
    \begin{subfigure}[b]{0.49\columnwidth}
        \vskip 0pt
        \includegraphics[height=3.35cm]{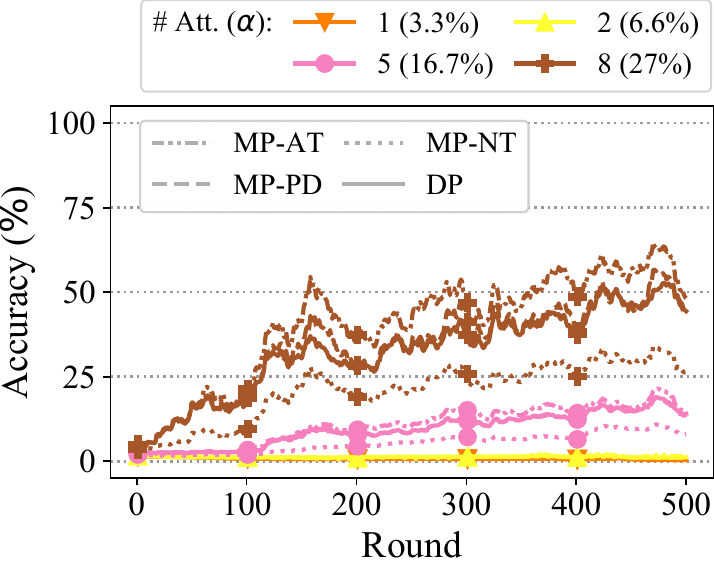}
        \caption{\taskIprototype}
        \lfig{analysis:increase_attackers:combined:fmnist}
    \end{subfigure}
    \begin{subfigure}[b]{0.49\columnwidth}
        \vskip 0pt
        \includegraphics[height=3.35cm]{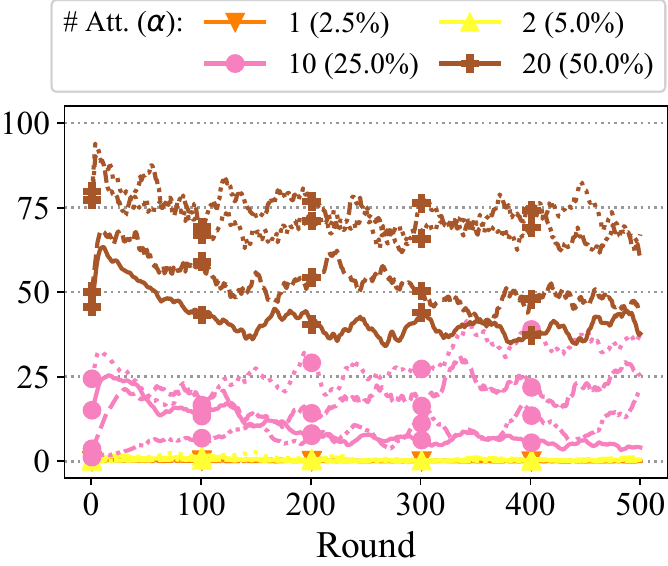}
        \caption{\taskIIprototype}
        \lfig{analysis:increase_attackers:combined:cifar}
    \end{subfigure}
    
    \caption{Comparison of model poisoning attacks for various \% of compromised clients per round. (M-$r$: 1.5)}
    \lfig{analysis:increase_attackers:all}
\end{figure}

\begin{figure}[t]
    \begin{subfigure}[b]{0.46\columnwidth}
        \vskip 0pt
        \includegraphics[height=3cm]{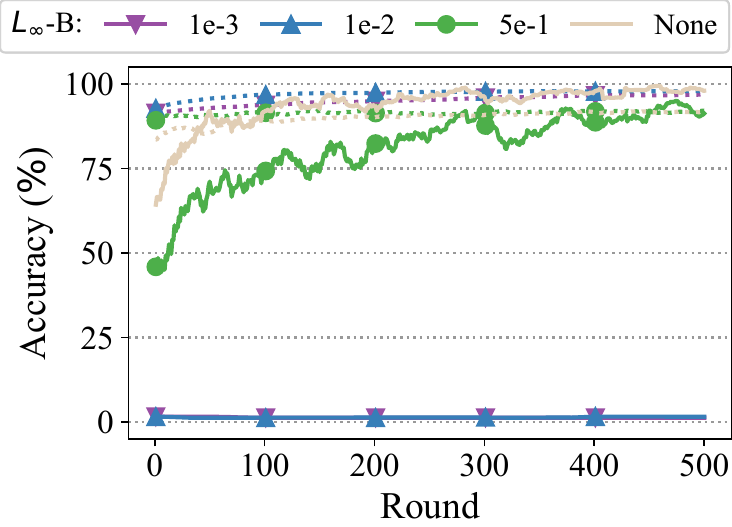}
        \caption{\taskIprototype}
        \lfig{apx:analysis:continuous_linf:fmnist}
    \end{subfigure}
    \hfill
    \begin{subfigure}[b]{0.5\columnwidth}
        \vskip 0pt
        \includegraphics[height=3cm]{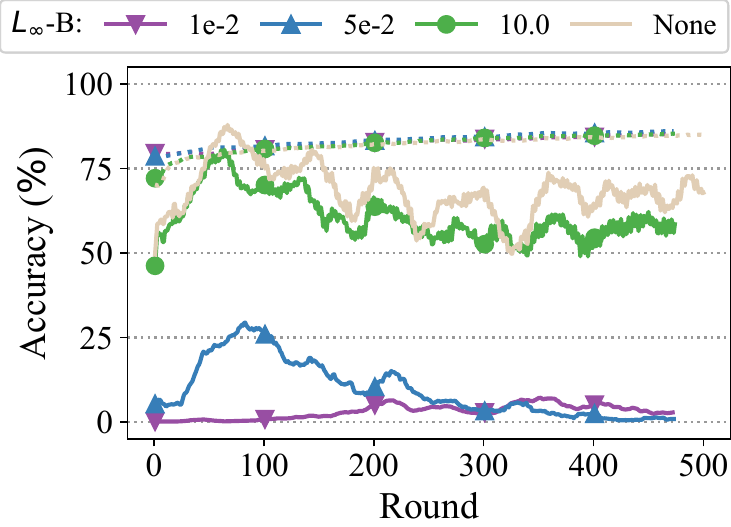}
        \caption{\taskIIprototype}
        \lfig{apx:analysis:continuous_linf:cifar}
    \end{subfigure}

    \vspace{6pt}
    \caption{Comparison of PGD attack under various static $L_\infty$-norm bounds.}
    \lfig{apx:analysis:continuous_linf}
\end{figure}

\fakeparagraph{$L_\infty$-norm bound.}
we show that an $L_\infty$-norm bound has the ability to defend against prototypical backdoor attacks,
similar to the static and adaptive $L_2$-norm bounds shown in Section~\rsec{sec:analysis}.
An $L_\infty$-norm bound of $0.01$ for \taskI and $0.05$ for \taskII is sufficient to successfully defend against a backdoor attack on prototypical samples,
without impacting model convergence or accuracy (\rfig{apx:analysis:continuous_linf}).
Similar to the $L_2$-norm bound, when the norm bound is too tight, model convergence is impacted.
Conversely, when the bound is too loose, the attacker is able to successfully inject the prototypical backdoor into the model.

\section{}
\lsec{pc}

In this appendix, we discuss the security of the probabilistic checking optimization for $L_\infty$-norm constraints in practical model poisoning attacks and justify our choice of security parameters.

The security of probabilistic checking is determined by the probability that a client is not detected (i.e., the failure probability) by the server while
submitting a malicious update that violates the $L_\infty$-norm constraint.
Let $\mathbf{w}$ be a malicious update vector of length $\ell$ that violates the $L_\infty$-norm constraint, i.e., $||\mathbf{w}||_\infty > B$,
because $p_v \cdot \ell$ for $p_v \in (0,1]$ parameters in $\mathbf{w}$ are above $B$.
The server selects a fraction $p_c \in (0,1]$ of the parameters $\mathbf{w'}$ in $\mathbf{w}$ uniformly at random and checks whether $||\mathbf{w'}||_\infty > B$.
We can derive the probability of not detecting this malicious update vector by realizing this random selection by the server is
    equivalent to drawing $p_c \cdot \ell$ parameters without replacement.
    The probability of not detecting is then defined as the probability that all $p_c \cdot \ell$ draws are below-bound parameters
    for a pool in which $1 - (p_v \cdot \ell)$ out of $\ell$ parameters are below the bound.
    This probability $\Pr(X=k)$ follows the hypergeometric distribution for $k$ draws.
    Specifically, we are interested in the case when $k = p_c \cdot \ell$, which defines the failure probability as
	\[ \Pr(X = p_c \ell) = Hyp(p_c \ell~|~\ell, \ell \cdot (1-p_v), p_c \ell)\]
          	\[  = \frac{\binom{\ell (1-p_v)}{p_c \cdot \ell}\binom{\ell-\ell \cdot (1-p_v)}{p_c\ell-p_c\ell}}{\binom{\ell}{p_c\ell}} \]
           \[ = \frac{\binom{\ell (1-p_v)}{p_c\ell}}{\binom{\ell}{p_c\ell}}.\]
    \\
    The security of the probabilistic checking approach depends on the assumption of the fractions $p_v$ and $p_c$, where $p_c$ can be set by the server.
    To configure $p_c$, \oursystem assumes a fixed minimal $p_v$ and a bounded failure probability of $1e-8$.
    In \rfig{fig:design:hypergeometric}, we present the number of parameters the server has to check for a fixed detection failure probability $\delta$ and fraction $p_v$ of parameters assumed to be above the bound.
    We argue that in practical model poising attacks that depend on scaling, a significant fraction (i.e., $\geq p_v$) of parameters have to be above the bound for an attack to be successful.
    Here, we analyze the fractions of parameters above the bound for generic attacks from \rsec{sec:analysis},
    and then show results for an adaptive attack in which the attacker optimizes for the number of scaled parameters.

\fakeparagraph{Generic Attacks.}
\rfig{apx:analysis:weight_distribution} shows the update weight distribution of a benign and a malicious update for the \taskIprototype single-shot attack.
For the malicious update, $81.8\%$ of parameters lie outside the $L_\infty$-bound of $0.01$, versus $0.87\%$ for the benign update.
In the continuous attack scenario, \taskIprototype, a similarly large fraction of parameters greater than $47\%$ are above the bound in each round (\rfig{apx:analysis:weight_distribution_overtime}).

\fakeparagraph{Adaptive attacks.}
In the adaptive case, we consider an adversary that optimizes for a minimal number of parameters above the bound in the poisoned update.
The adaptive attacker optimizes their model poisoning update by selecting a fraction of $K$ weights to scale based on their magnitude.
We instantiate the following \emph{top-$K$} adaptive attacker:
Upon receiving the global model, the attacker crafts a malicious update as in the generic case, but
scales only the $K \cdot \ell$ largest weights, clipping the remaining weights to the bound before sending the update to the server.
\rfig{apx:analysis:pcheck_security:topk} shows the performance of the top-$K$ \taskIIprototype single-shot attack for varying percentages of $K$.
Attacks are always successful for $K \ge 10\%$, which means the attacker does not have to scale the smallest $90\%$ of weights to attack successfully.
However, attack success starts to decrease for $K<10\%$, with success below $50\%$ for $K=5\%$.
Finally, the attack becomes completely ineffective ($0\%$ success) for $K \leq 2\%$.
In \taskIprototype, the attacker has to scale less update parameters for success, but still a significant fraction of over 2 percent must be above the bound.
In \rfig{apx:analysis:pcheck_real_vector}, we show the effect of varying $p_c$ on the detection rate for specific top-$K$ attack updates in both \taskIprototype and \taskIIprototype tasks.
The detection rate quickly converges to 1.0 as $p_c$ increases.
In our analysis, we opt for $0.005$ as a conservative estimate for $p_v$, which is well below the observations in the generic and adaptive attacks.

\begin{figure}[t]
    \begin{subfigure}[b]{0.49\columnwidth}
        \vskip 0pt
        \includegraphics[height=2.18cm]{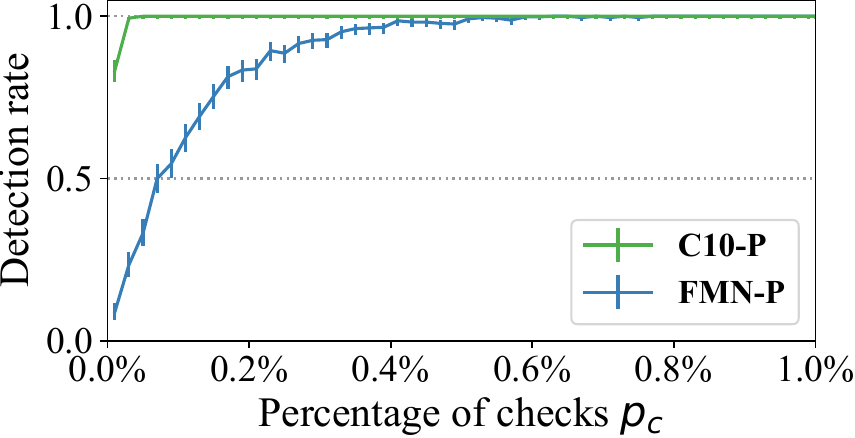}
        \caption{}
        \lfig{apx:analysis:pcheck_real_vector}
    \end{subfigure}
    \hfill
    \begin{subfigure}[b]{0.5\columnwidth}
        \vskip 0pt
        \includegraphics[height=2.18cm]{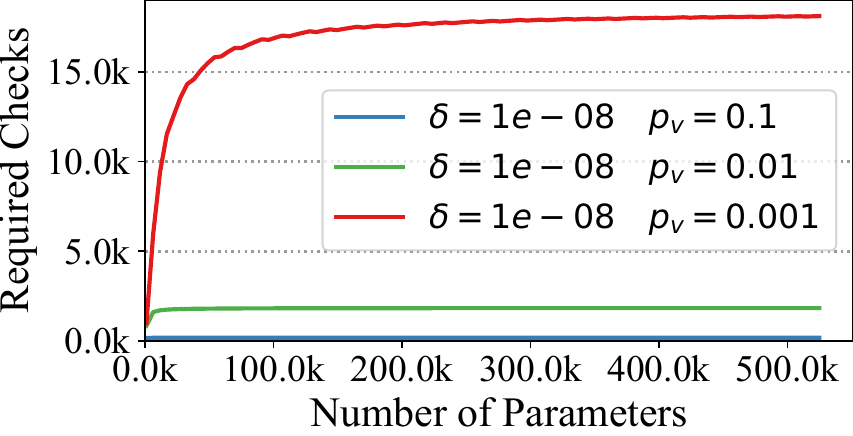}
        \caption{}
        \lfig{fig:design:hypergeometric}
    \end{subfigure}
    \caption{(a) The server only has to check a small fraction of parameters $p_c$ to ensure a high detection rate
    on two malicious updates created using the \emph{top-K} adaptive attack.
        (b) Number of update parameter ranges the server has to check with a failure probability of $\delta$ and the assumption that a fraction $p_v$ of the parameters are above the bound.}
\end{figure}

\begin{figure}[t]
    \begin{subfigure}[b]{0.49\columnwidth}
        \vskip 0pt
        \includegraphics[height=3cm]{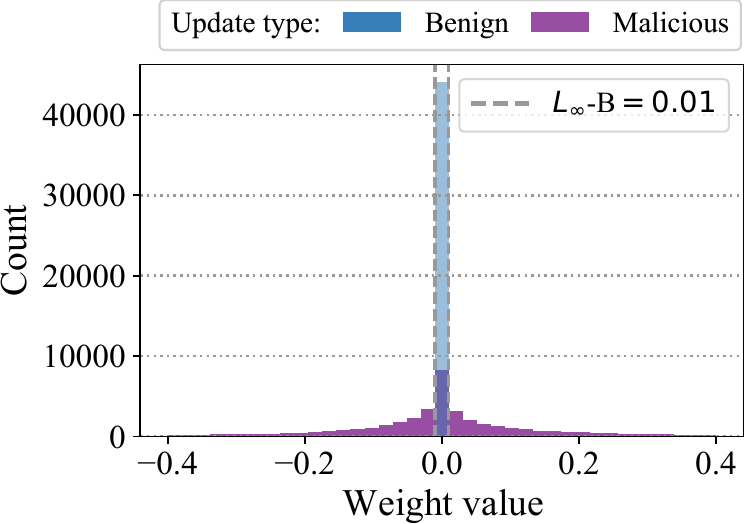}
        \caption{}
        \lfig{apx:analysis:weight_distribution}
    \end{subfigure}
    \hfill
    \begin{subfigure}[b]{0.47\columnwidth}
        \vskip 0pt
        \includegraphics[height=3cm]{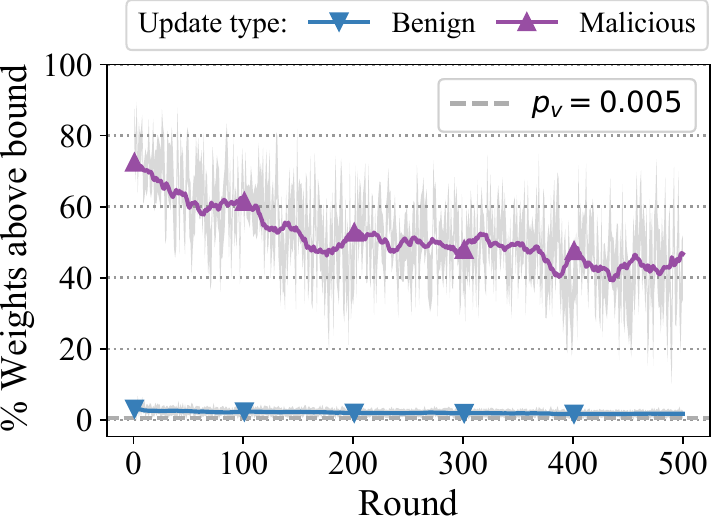}
        \caption{}
        \lfig{apx:analysis:weight_distribution_overtime}
    \end{subfigure}
    \caption{Generic attacks: (a) Weight distribution of a malicious update (\taskI, \taskIprototype).
    Parameters outside of the interval $[-0.4, 0.4]$ are hidden. (b) Percentage of weights outside the $L_\infty$-norm bound of 0.01 for malicious and benign clients per round (\taskI, \taskIprototype).
    The total number of weights is 44426. We plot the assumption $p_v = 0.005$ as used in the evaluation.}
    \lfig{apx:analysis:pcheck_weight_dist}
\end{figure}

\begin{figure}[t]
    \centering
    \vskip 0pt
    \includegraphics[width=0.35\textwidth]{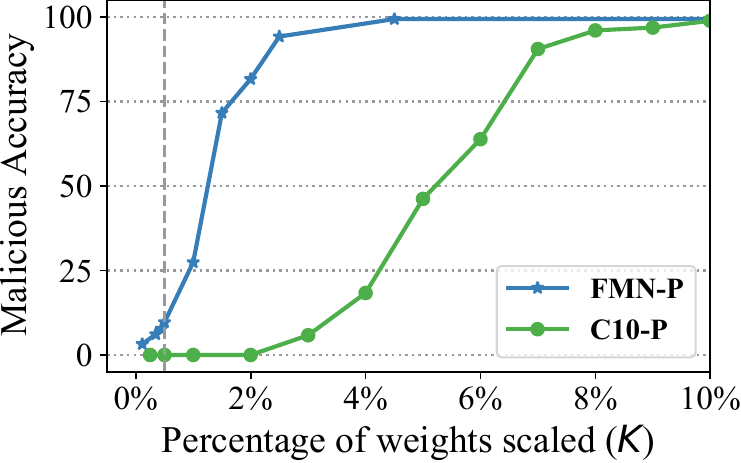}
    \caption{Top-$K$ success attack on probabilistic checking.}
    \lfig{apx:analysis:pcheck_security:topk}
    \vspace{-1em}
\end{figure}

\section{}
\lsec{apx:experimental_setup}
\vspace{-0.25em}
\subsection{Analysis Setup}
We first describe the details of the tasks in the analysis, and then continue with the configurations of the attacks for these tasks.
\fakeparagraph{Analysis Tasks.}
We consider two image classification tasks that have been used before in the context of \FL backdoor attacks to provide comparison to previous work.
The first task (\taskI) is a digit classification task on the Federated-MNIST dataset, with $n=3383$ available clients and $m=30$ clients selected per learning round, following~\cite{clipping-sun2019can,edge-case-backdoor}.
This dataset consists of samples from 3383 hand writers and is inherently non-IID because of the per-writer grouping, which simulates a natural distribution of training data.
The second task (\taskII) is an image classification on the CIFAR-10 dataset.
We divide the dataset among $n=100$ clients and select $m=40$ clients in each round.
We distribute the images to clients in a non-IID fashion using a Dirichlet distribution with parameter 0.9 in all dimensions.
The full hyperparameters for the setups of \taskI and \taskII are shown in \rtab{tab:hyperparams_benign}.
For \taskII, we use standard data augmentations (random horizontal flips and shifts) to reduce overfitting on the training dataset.

The global learning rate $\eta$ controls the fraction of the model that is updated each round.
For \taskI, the full model is updated with the average of $m=30$ client updates ($\eta = \frac{3383}{30}$).
For \taskII, we follow~\cite{Bagdasaryan2018-yx} and use a global learning rate of $1$, indicating that $40\%$ of the model is replaced every round.
We pre-train the models for both tasks once and use the same weights as a starting point for each experiment, similar to~\cite{edge-case-backdoor}.
The pre-trained models have not yet fully converged, to show the effect of different configurations on model convergence.
For \taskI, we use a pre-trained LeNet5 with 88\% test accuracy and for \taskII we use a ResNet-20 with 80\% test accuracy.
The full configuration of the models are given in \rsec{apx:ssec:modelarch}.
All experiments in our analysis, except those in \rfigs{analysis:singleshot}{analysis:single_shot_outlier}, show the accuracy as a moving average with a window size of 20,
to improve readability.

\fakeparagraph{Attack Tasks.}
For the \taskI task, we consider a backdoor attack targeting prototypical inputs that classifies images containing the number 7 from 30 randomly selected hand-writers as the number one instead (\taskIprototype, 310 images in total), following~\cite{clipping-sun2019can}.
For a \ourTailTarget backdoor attack, we follow~\cite{edge-case-backdoor} and use images of sevens from the ARDIS dataset that are mostly in the European style (with a horizontal bar in the middle), unlike the sevens in F-MNIST (\taskIedge, 660 images).
On the \taskII task, we use a prototypical backdoor attack from~\cite{Bagdasaryan2018-yx}: classifying images of green cars as birds (\taskIIprototype, 30 images).
For the \ourTailTarget backdoor attack, we use images from airplanes of Southwest airline that are collected by~\cite{edge-case-backdoor} to be introduced as \ourTailTarget backdoor images for CIFAR-10 (\taskIIedge, 784 images).

\fakeparagraph{Attack Success.}
For the \taskIprototype attack, attacker success is defined by the malicious classification accuracy on the backdoor images.
For the \taskIIprototype attack, we measure accuracy using three backdoor samples that were not made available to the attacker during training.
Test images are augmented into 200 versions per test image and used to evaluate the attacker success in the global model~\cite{Bagdasaryan2018-yx}.
For the \ourTailTarget attacks (\taskIedge, \taskIIedge), the images are already grouped into  train and test sets, the latter of which are used \mbox{to evaluate accuracy}.

\renewcommand{\arraystretch}{1.2}

\begin{table}[t!]
    \small
    \centering
    \resizebox{\columnwidth}{!}{%
        \begin{tabular}{ p{4.3cm} c c c }
            \hline
            \textbf{Client} & \taskI && \taskII \\
            \toprule
            \# Training samples & 341873 && 50000 \\
            Model & LeNet5 && ResNet-20 \\
            \# Classes & 10 && 10 \\
            \# Total number of clients & 3383 && 100 \\
            \# Selected clients per round & 30 && 40 \\
            \# Local epochs per round & 5 && 2 \\
            Batch size & 32 && 64 \\
            Optimizer & \multicolumn{3}{c}{SGD} \\
            Learning rate & 0.01 && 0.02 \\
            \hline
        \end{tabular}}
    \caption{Hyperparameters for honest clients.}
    \ltab{tab:hyperparams_benign}
    \vspace{-1em}
\end{table}
\renewcommand{\arraystretch}{1}

\fakeparagraph{Attack Strategies.}
We discuss the configuration of the backdoor injection attacks used by the adversary to create a malicious model update.
We explore two attacker models: data poisoning and model poisoning.
In data poisoning strategy, the attacker follows the training process (i.e., hyperparameters) as the benign clients,
but is allowed to insert a number of malicious samples into each batch according to the poison ratio.
For model poisoning, we explore three different attacks that are adaptive with regards to the norm bound that is enforced by the server.

The model poisoning clients train the model for a fixed amount of SGD steps, determined by the number of epochs and batches shown in~\rtab{tab:hyperparams_attacker}.
Each batch contains benign samples, and a number of target samples mislabeled with the target class.
The amount of malicious samples in each batch is controlled by the poison ratio.
The optimal backdoor injection attack also includes benign samples because it allows the attacker to move the model into a direction where both objectives are satisfied simultaneously, making the attacker's update harder to reverse by honest client updates~\cite{Bagdasaryan2018-yx,edge-case-backdoor}.
We apply learning rate decay to malicious training to ensure the model update has a good accuracy on both the main and backdoor tasks. %
The learning rate starts at 0.1 and decays step-wise with $\frac{1}{10}$ every $\frac{1}{3}$ of the total number of steps.
We now discuss details specific to the attacks.

\fakeparagraph{PGD}
When an $L_p$-norm bound $B$ is present, the adversary uses Projected Gradient Descent (PGD) to adaptively craft malicious updates.
Concretely, after every SGD step, the adverary projects model update $\Delta\hat{\mathbf{w}}$ back to
a point in the constraint set.
For the norm bound, this set is defined as $||\Delta\hat{\mathbf{w}}||_p \leq \frac{B}{\gamma}$, where $\gamma$ is the scaling factor.
More precisely, similar to~\cite{edge-case-backdoor}, during an epoch we project the update back onto a slightly larger space ($1.2\frac{B}{\gamma}$).
Only after the last SGD step of the epoch do we use $\gamma$ as a scaling factor to satisfy the norm constraint (\rtab{tab:hyperparams_attacker}).
In practice, this approach leads to a higher attacker success, similar as to what was reported in~\cite{edge-case-backdoor}.

\fakeparagraph{Neurotoxin}
We choose a mask ratio $k=5$ which means that the malicious update is projected onto the bottom $95\%$ of frequently updated weights using PGD.
This value for the mask ratio achieved good overall performance in both the original attack paper and in our results.

\fakeparagraph{Anticipate}
We follow the hyperparameters used in the original attack as closely as possible and set the number of anticipate steps $k=9$ for \taskII.
We choose $k=5$ anticipate steps for \taskI because Federated-MNIST dataset was not used to evaluate the attack.

\renewcommand{\arraystretch}{1.2}
\begin{table}[t!]
    \small
    \centering
    \resizebox{\columnwidth}{!}{%
        \begin{tabular}{p{5.5cm} c c c}
            \hline
            \textbf{Setup} & \taskI && \taskII \\
            \toprule
            \# Local epochs per round \qquad \qquad & 10  && 6  \\
            \# Batches & 25 && 10 \\
            Batch size & 32 && 64 \\
            \# Poison samples per batch & 12 && 20 \\
            Optimizer & \multicolumn{3}{c}{SGD} \\
            Learning rate & 0.1 && 0.1 \\
            Scale factor & 30 && 100 \\
            \hline
        \end{tabular} }
    \caption{Hyperparameters for model poisoning attacks}
    \ltab{tab:hyperparams_attacker}
    \vspace{-1em}
\end{table}
\renewcommand{\arraystretch}{1}

\vspace{-0.75em}
\subsection{\oursystem Evaluation Setup}
\vspace{-0.5em}
We now describe additional details of the experiments performed as part of the evaluation.

\fakeparagraph{Microbenchmarks.}
For each reported result, we take the average of 4 measurements recorded after a single warm-up mock execution.

\fakeparagraph{End-to-end.}
We evaluate the end-to-end performance on four different models.
We use the two models that are also used in analysis and we evaluate on two additional models.
ResNet-20~\cite{resnet1} (\evalCIFARL, 273k params) and LeNet5~\cite{lenet5} (\evalCIFARS, 62k params).
In addition, we evaluate the performance for a smaller CNN network (\evalMNIST, 19k params) and a text-generative LSTM model (\evalShakespeare, 818k params).
An overview of the tasks and models is shown in \rtab{tab:evaluation_tasks}.
The configuration of the end-to-end setup experiments are similar to those of the analysis,
except for some minor changes.
We increase the number of clients in the evaluation to 48 to benchmark the performance of a larger deployment.
In each training round, all 48 clients are selected.
For \evalMNIST, we use a learning rate of 0.05, for \evalCIFARS a learning rate of 0.01, for \evalCIFARL a learning rate of 0.05 and for \evalShakespeare 0.3.
In \oursystem, we apply 8-bit probabilistic quantization with 7 fractional bits for the floating point numbers in the client \mbox{updates for encryption.}

\vspace{-0.5em}
\subsubsection{Model Architectures}
\lsec{apx:ssec:modelarch}
We use the following model architectures for analysis and evaluation (end-to-end) experiments.

\fakeparagraph{CNN.}
For the \evalMNIST task, we use a standard convolutional neural network (CNN) architecture,
consisting of two convolution layers of 8 and 4 filters, respectively, with kernel size 3-by-3.
These are followed by a 2-by-2 max-pooling layer, a fully connected layer of 32 units and then the output layer.
The model has 19k trainable parameters.

\fakeparagraph{LeNet5.}
We make use of the standard LeNet5~\cite{lenet5} convolutional neural network for both \taskI in the analysis
and \evalCIFARS.
Note that the input layers for both applications of the model are different: 28-by-28, 1 channel for \taskI and 32-by-32, 3 channels for \evalCIFARS.

\fakeparagraph{ResNet-20.}
We use the ResNet~\cite{resnet1} model for experiments in the analysis (\taskI) as well as in evaluation experiments (\evalCIFARL).
We use the ResNet implementation provided by the Keras ResNet CIFAR-10 example\footnote{\url{https://github.com/rstudio/keras/blob/master/vignettes/examples/cifar10_resnet.py}},
corresponding to the ResNet20v1 model.

\fakeparagraph{LSTM.}
We use the LSTM architecture as used in the LEAF federated learning benchmark~\cite{federated-mnist} and the original federated learning paper~\cite{McMahan2017-gd}.
The architecture consists of an embedding layer of size 8 followed by a two-layer stacked LSTM with 256 hidden units.
The LSTMs are connected to dense output layer with width 80, which corresponds to the vocabulary size.

	}
}{
	\bibliography{biblio}

	\appendices
	
}

\end{document}